\documentclass[pre,showpacs,superscriptaddress,amsfonts,amsmath,twocolumn,floatfix]{revtex4}
\usepackage[sort&compress]{natbib}
\usepackage[final]{graphicx}
\usepackage{color}
\def\kap{\ensuremath{\boldsymbol\kappa}}
\newcommand{\vct}[1]{\mathbf{#1}}

\def\Pe{\ensuremath{\text{\textit{Pe}}}}
\def\kT{\ensuremath{k_\text{B}T}}
\DeclareMathOperator\smol{\Omega}
\DeclareMathOperator\dsmol{\delta\Omega}
\DeclareMathOperator\proj{{\mathcal P}}
\DeclareMathOperator\projQ{{\mathcal Q}}

\begin{document}
\title{From Equilibrium to Steady-State Dynamics after Switch-On of Shear}
\date{\today}
\def\unikn{\affiliation{Fachbereich Physik, Universit\"at Konstanz,
  78457 Konstanz, Germany}}
\def\zk{\affiliation{Zukunftskolleg, Universit\"at Konstanz,
  78457 Konstanz, Germany}}
\def\dlr{\affiliation{Institut f\"ur Materialphysik im Weltraum,
  Deutsches Zentrum f\"ur Luft- und Raumfahrt (DLR), 51170 K\"oln, Germany}}
\author{Matthias Kr\"uger}\unikn
\author{Fabian Weysser}\unikn
\author{Thomas Voigtmann}\unikn\dlr\zk

\begin{abstract}
A relation between equilibrium, steady-state, and waiting-time dependent
dynamical two-time correlation functions in dense glass-forming liquids
subject to homogeneous steady shear flow is discussed. The systems under study
show pronounced shear thinning, i.e., a significant speedup in their
steady-state slow relaxation as compared to equilibrium. An approximate
relation that recovers the exact limit for small waiting times is derived
following the integration through transients (ITT) approach for the
nonequilibrium Smoluchowski dynamics, and is
exemplified within a schematic model in the framework of the
mode-coupling theory of the glass transition (MCT).
Computer simulation results for the tagged-particle density correlation
functions corresponding to wave vectors in the shear-gradient directions
from both event-driven stochastic dynamics of a
two-dimensional hard-disk system and from previously published
Newtonian-dynamics simulations of a three-dimensional soft-sphere mixture
are analyzed and compared with the predictions of the ITT-based approximation.
Good qualitative and semi-quantitative agreement is found. 
Furthermore, for short waiting
times, the theoretical description of the waiting time dependence shows
excellent quantitative agreement to the simulations. This confirms the accuracy of the central approximation used
earlier to derive fluctuation dissipation ratios (Phys. Rev. Lett. {\bf 102}, 135701). For intermediate waiting
times, the correlation functions decay faster at long times than the stationary
ones. This behavior is predicted by our theory and observed in simulations. 
\end{abstract}

\pacs{82.70.Dd,
64.70.P-,
05.70.Ln,
83.60.Df
}
\keywords{Shear,Diffusion}

\maketitle

\section{Introduction} 

The application of shear flow to dense liquids can dramatically change their
transport and relaxation processes. Even if the timescale set by the applied
shear rate $\dot\gamma$ is slow compared to a typical single-particle relaxation time $\tau_0$
(i.e., the P\'eclet number $\Pe_0=\dot\gamma\tau_0\ll1$), it can interfere with and supersede
the slow relaxation times $\tau_\alpha$ of the system (i.e., the ``dressed'' P\'eclet,
or Weissenberg number $\Pe=\dot\gamma\tau_\alpha\gg1$). In this case, the slow relaxation of the
system is usually found to be accelerated by the shear flow, a phenomenon
known from colloidal suspensions as shear thinning, because a pronounced
decrease in the apparent viscosity results. To this decrease corresponds
an increase in the single-particle diffusivities. These changes in
transport processes are found even if the average static structure of the
system (at least as measured through two-point correlation functions)
changes only slightly.

The drastic change from equilibrium to steady-state transport
properties begs the question about transient dynamics: what happens if such
a slowly relaxing liquid is suddenly subjected to shear, regarding its dynamical
correlations
as the system progresses from equilibrium towards
its steady state? One can investigate these effects most easily by looking at
the waiting-time dependent dynamical two-time correlation functions: switching
on the external shear flow at $t=0$,
one measures the correlations of
dynamical variables between some waiting time $t_w>0$, and a correlation
time $t=t_w+\tau>t_w$. Of particular interest are the so-called
transient correlation functions, obtained for $t_w=0$, to be compared with
the reference cases of equilibrium and steady state.

In the present contribution, we address this issue by presenting an
approximate relationship between these three relevant types of dynamical
two-point correlation functions.

Recently, the dynamical evolution after switching on shear flow of steady
rate $\dot\gamma$ has been addressed \cite{Zausch08} by a combination of
techniques: theoretically, in the framework of
mode-coupling theory (MCT) for colloidal rheology and an integration-through
transients (ITT) approach, experimentally, using confocal microscopy, and
with computer simulation for a damped Newtonian-dynamics model. 
All three methods yield a consistent picture: a shear stress $\sigma(t)$
builds up at $t>0$ after switch-on, but does not grow monotonically towards
its steady-state value $\sigma_\infty$.
It rather exhibits an intermediate ``overshoot'' at times corresponding
to an overall strain $\gamma=\dot\gamma t\approx0.1$. Such stress overshoot
phenomena are in fact well-known \cite{Varnik04,Rottler03,Tanguy}, but despite their
ubiquity, their microscopic origin, in particular for the fully homogeneous
flow profiles studied, remains somewhat vague.
In Ref.~\cite{Zausch08}, simulations were able to connect it
to a sudden change in the mean-squared displacement (MSD) of a tracer
particle: even for the directions perpendicular to the shear flow
where no explicit advection occurs, one observes a
super-diffusive regime as the transient MSD, $\delta r^2(t,t_w=0)$,
leaves the equilibrium curve
around $\gamma\approx0.1$, to cross over to the (much larger) steady-state
curve which it reaches at $\gamma\approx1$. In this regime, MD simulations
found motion to be almost ballistic, $\delta r^2(t\approx0.1/\dot\gamma,0)
\approx t^x$ with $x\approx2$. In experiment (closer to Brownian dynamics),
this superdiffusion was not
as pronounced, yielding $x$ only slightly larger than $1$.
Within MCT-ITT and an additional ad-hoc approximation akin to a
generalized Stokes-Einstein relation, it could be shown that the stress
overshoot and superdiffusion are directly connected and originate from
an overrelation of microscopic stresses: the transient stress autocorrelation
function (called a dynamical shear modulus) does not decay monotonically
to zero, but exhibits a ``dip'' in the corresponding strain regime where
it becomes slightly negative just before reaching its zero long-time limit.

Thus, the details on the evolution from equilibrium dynamics to
far-from-equilibrium dynamics under shear are encoded in transient
correlation functions. This raises a two-fold interest in these transient
correlation functions: first, given information on both equilibrium and
the steady state, what can one infer about the transient dynamics?
Second, recalling that MCT and ITT generically build upon the transient
correlation functions, can one test their generic implications?

The relationship among the various two-point dynamical correlation functions
that we present in the following, builds upon the ITT formalism, without
explicit reference to MCT. It should thus hold quite generally, at least
qualitatively. We demonstrate this by comparing with computer-simulation
data for both a Brownian and a non-Brownian system.
For dense liquids and colloidal suspensions in equilibrium, it is
a well-tested paradigm, that the long-time behavior of the correlation
functions does not (up to an overall time unit) depend on the type of
short-time motion, be it ballistic (Newtonian dynamics) or diffusive (Brownian
colloidal particles). This in fact defines the regime of quiescent ``structural
relaxation'', where slow relaxation processes arising from collective
caging of particles govern the dynamics of the system.

This equivalence of Newtonian and Brownian systems does not need to hold
far from equilibrium, although similar shear-thinning
effects are seen both in colloidal suspensions and atomistic metallic
melts. Indeed, differences were observed regarding the extent of superdiffusive
motion in the transient MSD \cite{Zausch08}.
Note that in the MD simulations of Ref.~\cite{Zausch08},
shear was implemented through the boundaries of the (periodically repeated)
simulation box only, by Lees-Edwards boundary conditions. Hence,
particles at the center of the box remain at rest also for a short time
after switching on the flow, until a linear shear profile propagates from
the boundaries towards the center. It was argued that this time scale is
short compared to the $\dot\gamma t\approx0.1$ of interest. In
Brownian dynamics, the issue can be set aside, as there one immediately
modifies the solvent flow profile throughout the box when implementing shear
(in addition to Lees-Edwards boundary conditions). This situation also
being closer to what MCT-ITT models, is another motivation to complement the
MD simulation data already partially discussed in Ref.~\cite{Zausch08}
with new simulations incorporating diffusive short-time motion.

The peculiar features of the transient correlation functions not only
highlight possible differences in the non-equilibrium response of the
different system types. They also provide direct tests of the MCT-ITT
formalism. At the core of ITT is a reformulation of non-equilibrium
averages in terms of history integrals over equilibrium averages where
the full nonequilibrium time-dependence is kept in the evolution of the
dynamical variables. In the case of steady shear, this is precisely the
transient correlation function measurable in experiment or simulation where
a shear flow is switched on immediately at $t=0$.

While the transient ($t_w=0$) correlation function is the natural object to
be treated in MCT-ITT, steady-state and more generally $t_w$-dependent
correlation functions have, in this approach, to be calculated afterwards.
Our main result here enables MCT-ITT to do just that: by obtaining equilibrium
and transient correlation functions from any theory, correlators for all
$t_w$ can be expressed.

The paper is structured as follows: after some notational clarification
(Sec.~\ref{sec:corr}), we present in Sec.~\ref{sec:susc} the main theoretical
derivation of our formulas. Section~\ref{sec:schematic} is devoted to a
schematic-MCT illustration of the results, while Sec.~\ref{sec:compsim}
presents the analysis of computer-simulation data.
Section~\ref{sec:discuss} concludes the
discussion.

\section{Correlation functions}\label{sec:corr}

Given two dynamical fluctuating variables $\delta f$ and $\delta g$, i.e.,
functions with zero average that depend on
the state-point $\Gamma$ of the system,
one defines the two-point correlation function $C^{fg}(t,t_w)$ for $t\ge t_w$ \cite{Risken},
\begin{multline}\label{eq:corr}
  C^{fg}(t,t_w)=\iint d\Gamma d\Gamma'\,\delta g(\Gamma)
  P(\Gamma,t|\Gamma',t_w)\times\\ \delta f^*(\Gamma')\Psi_{t_w}(\Gamma')\,.
\end{multline}
Here, $P(\Gamma,t|\Gamma',t_w)$ denotes the conditional probability that the
system resides at state point $\Gamma$ at time $t=t_w+\tau$, given it was
at state point $\Gamma'$ at time $t_w$. $\Psi_{t_w}(\Gamma')$ is the
probability that the system is at $\Gamma'$ at time $t_w$, and of course
just the (nonequilibrium) distribution function. The latter is assumed to
be equal to the equilibrium distribution for times $t<0$,
$\Psi_{t<0}(\Gamma)=\Psi_e(\Gamma)$, and to asymptotically reach
a time-independent steady state for long times,
$\Psi_{t\to\infty}(\Gamma)=\Psi_s(\Gamma)$.
These reference limits define the equilibrium and
steady-state ensemble averages,
\begin{subequations}\label{eq:averages}
\begin{gather}
\langle\operatorname{\cdots}\rangle=\int d\Gamma \Psi_e(\Gamma) \cdots\,,\\
\langle\operatorname{\cdots}\rangle^{(\dot\gamma)}=\int d\Gamma \Psi_s(\Gamma) \cdots\,.
\end{gather}
\end{subequations}

\begin{figure}
\centering{\includegraphics[width=0.6\linewidth]{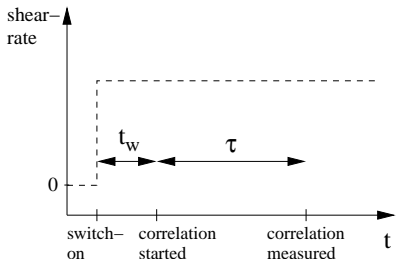}}
\caption{\label{fig:waiting}
  Schematic representation of the times appearing in the two-point
  correlation functions, Eq.~\eqref{eq:corr}: waiting time $t_w>0$,
  measurement time $t$, and correlation time $\tau=t-t_w>0$. $t=0$
  corresponds to the time where homogeneous, linear shear flow is
  instantaneously switched on.
}
\end{figure}

The conditional probability $P$ in Eq.~\eqref{eq:corr}
encodes the dynamics of the system, subject
to the external field.
We will use the following property \cite{Risken}:
for the case of switching on a constant shear flow,
$P(\Gamma,t_w+\tau|\Gamma',t_w)$ becomes independent on $t_w$ for all
$t_w>0$. 
For Brownian
dynamics, this means that both $\Psi_{t_w}$ and $P$ obey
the same differential (Smoluchowski) equation, i.e., the underlying stochastic
process is assumed to have a Markovian property \cite{vanKampen}.
As long as $\Psi_{t_w}(\Gamma')$ evolves, the correlation
function $C^{fg}(t,t_w)$ will depend on its two time arguments separately;
we will generally call these functions ``waiting-time dependent'' (for
waiting time $t_w$). When it is clear from the context, we omit the
superscript denoting the variables and abbreviate $C_{t_w}(\tau)\equiv
C(t_w+\tau,t_w)$. While potentially interesting, we ignore correlation
functions formed with $t_w<0$ and $t>0$.
Figure~\ref{fig:waiting} schematically summarizes the
sequence of correlation and measurement times.

As $t_w\to\infty$, the steady-state 
correlation function is approached, which we denote by
$C_\infty(\tau)\equiv C(t,+\infty)$.
Correspondingly, as $t_w<-\tau$,
one obtains the equilibrium correlation function,
$C_e(\tau)\equiv C(t,-\infty)$. Both $C_\infty(\tau)$ and
$C_e(\tau)$ are functions of the time difference $\tau$ only.
Among the general waiting-time dependent correlation functions, a particular
role is played by the $t_w=0$ case: recalling $\Psi_0=\Psi_e$, we recognize the
so-called transient correlation function $C_0(\tau)\equiv C(t,0)$,
where the time evolution as determined by the transition rates $P$ is
the nonequilibrium one, but averaging is performed with the
equilibrium distribution.

Note that we do not in general assume $C_e(\tau)$ to decay to zero
for $\tau\to\infty$: in (idealized) glass states, it attains a finite
positive long-time limit, called the nonergodicity factor or glass form
factor \cite{Goetze91}.

In comparing with computer simulations, we set for simplicity $f=g$ to be
the one-particle microscopic number density, $\delta f=\exp[i\vct q\cdot\vct r_s]$
with wave vector $\vct q$ and the position of the singled-out particle
$\vct r_s$. To further simplify the discussion, we restrict ourselves
to wave vectors perpendicular to the flow direction. This obliterates the
need of introducing wave-vector advection in order to account for the
affine motion imposed by the shear \cite{Fuchs08}.

Connected to the zero-wavevector limit of tagged-particle density fluctuations
is of course the mean-squared displacement (MSD),
\begin{multline}
  \delta r^2(t,t_w)=\iint d\Gamma d\Gamma'\,
  [\vct r_s(\Gamma)-\vct r_s(\Gamma')]^2\times\\
  P(\Gamma,t|\Gamma',t_w) \Psi_{t_w}(\Gamma')\,,
\end{multline}
also schematically written as
$\delta r^2(t,t_w)=\langle[\vct r_s(t)-\vct r_s(t_w)]^2\rangle^{si}$
where $\langle\operatorname\cdots\rangle^{si}$ denotes averaging over particles and runs as done in a simulation.
Again, the MSD comes in its equilibrium ($\delta r^2_e(\tau)$),
steady-state ($\delta r^2_\infty(\tau)$), and transient
($\delta r^2_0(\tau)$) varieties.

We generally consider a system of $N$ spherical particles without internal
degrees of freedom, enclosed in a volume $V$. Choosing units, we set
the thermal energy $\kT=1$ throughout. For the stochastic-dynamics simulations
as well as the theory, we assume diffusive short-time motion governed by
a bare diffusion coefficient $D_0=1$; we do not take into account explicit
solvent or hydrodynamic interactions among the particles.
A typical interaction diameter of the particles, $\sigma=1$, sets
the unit length.

We choose coordinates such that the external flow acts in the Cartesian
$x$-direction (called flow direction), and varies along $y$ (called gradient
direction). In 3D, the system is invariant along the $z$-axis (neutral direction). Thus, the velocity field induced by the shear can be written
as $\vct v(\vct r)=\dot\gamma y\hat{\vct x}$ with the velocity-gradient
tensor $\kap=\dot\gamma\hat{\vct x}\hat{\vct y}$, where $\hat{\vct x}$
is a unit vector in the direction of $\vct x$.

\section{Microscopic Theory}\label{sec:susc}

Within the ITT formalism, we now derive expressions for the time-dependent
correlation functions of interest. We begin by recalling the exact
starting points of ITT, before introducing approximations that lead to
our final result, presented in Sec.~\ref{sec:mainresult}.

\subsection{Integration Through Transients}

For the theoretical derivation, assume interaction forces among the particles
to be $\vct F_i=-\vct\partial_iU$ ($i=1,\ldots N$), where $U$ is the total
potential energy of the system. In the thermodynamic limit, the particle
distribution function $\Psi_t(\Gamma)$ of the Brownian system
subject to homogeneous shear flow described by the velocity-gradient
tensor $\kap(t)$,
is then taken to obey the Smoluchowski equation \cite{Dhont,Fuchs05},
\begin{subequations}\label{eq:smol}
\begin{gather}
\partial_t \Psi_t(\Gamma)=\smol(t)\Psi_t(\Gamma)\,,\\
\smol(t)=\smol_e+\dsmol(t)
  =\sum_{i}\boldsymbol{\partial}_i\cdot\left[
  \boldsymbol{\partial}_i-{\bf F}_i - \kap(t)\cdot\vct{r}_i\right]\,.
\end{gather}
\end{subequations}
Here, $\smol$ is the Smoluchowski operator (SO), consisting of the
equilibrium (quiescent) contribution,
$\smol_e=\sum_{i}\boldsymbol{\partial}_i
\cdot[\boldsymbol{\partial}_i-{\bf F}_i]$, and the nonequilibrium term
representing homogeneous driving.
For the case considered here (switching on constant shear flow of rate
$\dot\gamma$ at $t=0$),
$\dsmol(t)=\dsmol=-\sum_i\boldsymbol{\partial}_i\cdot \kap\cdot\vct{r}_i$
independent on $t$ for $t>0$ and zero else. Hence,
\begin{equation}
  \smol(t)=\begin{cases} \smol_e & \text{for $t<0$,}\\
                          \smol^{(\dot\gamma)} & \text{for $t>0$,}
  \end{cases}
\end{equation}
where $\smol^{(\dot\gamma)}$ does not depend on time.
The equilibrium distribution function $\Psi_e$ is the stationary solution
of Eq.~\eqref{eq:smol} without shear, $\smol_e\Psi_e\equiv0$, viz.
$\Psi_e\propto\exp(-U/\kT)$. Including shear,
$\smol^{(\dot\gamma)}\Psi_s\equiv0$ defines the steady-state
distribution.
In this stationary state, the distribution function is time-shift invariant
as in equilibrium, but the system is not in thermal equilibrium due to
a non-vanishing probability current \cite{Krueger09}.

The integration through transients (ITT) formalism allows to reformulate
the nonequilibrium averages formed with the (unknown) $t_w$-dependent
distribution $\Psi_{t_w}$ in terms of equilibrium averages.
Formally solving the Smoluchowski equation, Eq.~\eqref{eq:smol}, as an
integral equation with the boundary condition $\Psi(t=0)=\Psi_e$, one gets
for $t_w\ge0$,
\begin{equation}
\Psi(t_w)=e^{\smol^{(\dot\gamma)} t_w} \Psi_e= \Psi_e+\int_0^{t_w}ds\,
\smol^{(\dot\gamma)} e^{\smol^{(\dot\gamma)} s} \Psi_e\,.\label{eq:notc}
\end{equation}
Recalling $\smol^{(\dot\gamma)}\Psi_e=\dsmol\Psi_e=\sigma_{xy}\Psi_e$ \cite{Fuchs05}, integration by parts
yields
\begin{equation}
\int d\Gamma \Psi_{t_w} \operatorname{\cdots}
  =\int d\Gamma \Psi_e\left[1+\dot\gamma\int_0^{t_w} ds\,
   \sigma_{xy} e^{\smol^\dagger s}\right] \operatorname{\cdots}\label{eq:itttw}
\end{equation}
with the microscopic (potential) stress tensor element
$\sigma_{xy}=-\sum_iF_i^x y_i$ \cite{Fuchs05}. Here,
$\smol^\dagger=\sum_i[\boldsymbol{\partial}_i+\vct{F}_i+\vct{r}_i\cdot\kap^T]
\cdot\boldsymbol{\partial}_i$ is the operator adjoint to
$\smol^{(\dot\gamma)}$.
With $t_w\to\infty$, the steady-state average in
Eq.~\eqref{eq:averages} immediately follows 
\begin{equation}
\langle\operatorname\cdots\rangle^{(\dot\gamma)}
=\langle\operatorname\cdots\rangle
+\dot\gamma\int_0^{\infty} ds\,
\langle\sigma_{xy} e^{\smol^\dagger s} \operatorname\cdots\rangle
\label{eq:itt}.
\end{equation}

The Smoluchowski equation, Eq.~\eqref{eq:smol}, is also taken
to determine the conditional probability $P$ appearing in
Eq.~\eqref{eq:corr}, and thus \cite{Fuchs05,Risken},
for $t_w>0$,
$P(\Gamma,t_w+\tau|\Gamma',t_w)=\exp[\smol^{(\dot\gamma)}\tau]
\delta(\Gamma-\Gamma')$.
Rewriting in terms of the adjoint operator, this gives
\begin{subequations}
\begin{equation}
  C_\infty^{fg}(\tau)=\left\langle\delta f^*
  e^{\smol^\dagger \tau}\delta g\right\rangle^{(\dot\gamma)}
\end{equation}
and, for the transient correlation function,
\begin{equation}\label{eq:tran}
  C_0^{fg}(\tau)=\left\langle\delta f^*
  e^{\smol^\dagger \tau}\delta g\right\rangle\,.
\end{equation}
Note its distinction from the equilibrium correlation function,
where the equilibrium adjoint SO, $\smol_e^\dagger$, appears,
\begin{equation}
  C_e^{fg}(\tau)=\left\langle\delta f^*
  e^{\smol_e^\dagger \tau}\delta g\right\rangle\,.
\end{equation}
\end{subequations}
For the general two-time correlation function at finite $t_w>0$,
inserting into Eq.~\eqref{eq:itttw} gives
\begin{multline}
  C_{t_w}^{fg}(\tau)=C_0^{fg}(\tau)\\
  +\dot\gamma\int_0^{t_w} ds\,
  \left\langle\sigma_{xy}e^{\smol^\dagger s}\delta f^*
  e^{\smol^\dagger \tau}\delta g\right\rangle.\label{eq:2ti}
\end{multline}
This equation is formally exact, although the evaluation of the
dynamical three-point average in the integral will generally be hard.
We are therefore forced to introduce approximations at this point.

\subsection{Approximations for Correlation Functions}

To simplify the discussion, we now restrict ourselves 
to auto-correlation functions ($\delta f=\delta g$) of dynamical
fluctuations without explicit shear-advection, $\delta f\equiv f(\{y_i,z_i\})$.
Similar results can be expected for correlation functions involving
shear-advected quantities (i.e., dynamical variables explicitly depending
also on positions $x_i$ along the shear flow), but one then has to be
careful in first extracting the affine transformations induced by the
steady shear.

To obtain a tractable expression for the general $t_w$-dependent correlation
function, let us apply a familiar identity in
the Zwanzig-Mori operator formalism
(cf. Eq.~(11) in Ref.~\cite{Fuchs02b} and also Ref.~\cite{Goetze89}):
introducing a projector onto $\delta f$,
$\proj_f=\delta f\rangle\langle\delta f^*\delta f\rangle^{-1}\langle\delta f^*$,
with complement $\projQ_f=1-\proj_f$, we get from Eq.~\eqref{eq:2ti},
\begin{multline}\label{eq:ZM}
  C_{t_w}^f(\tau)=C_0^f(\tau)\left[1+\dot\gamma\int_0^{t_w}ds\,\frac{
  \left\langle\sigma_{xy}e^{\smol^\dagger s}\delta f^*\delta f\right\rangle}
  {\langle\delta f^*\delta f\rangle}\right]\\
  +\dot\gamma\int_0^{\tau} d\tau'\int_0^{t_w}ds\,\frac{
  \left\langle\sigma_{xy}e^{\smol^\dagger s}\delta f^*{\mathcal U}(\tau-\tau')
  \delta f\right\rangle}{\langle\delta f^*\delta f\rangle}
  C_0^f(\tau')\,,
\end{multline}
with the restricted time evolution operator
\begin{equation}
  {\mathcal U}(a)=\projQ_f\exp[\projQ_f\smol^\dagger\projQ_f a]\projQ_f
  \smol^\dagger\,.
\end{equation}

We thus identify two contributions to the difference between the
nonequilibrium waiting-time dependent correlator and the transient one.
The first is a $\tau$-independent renormalization of the equal-time value and
corresponds, e.g.,
to the difference of distorted and equilibrium static structure factor
if $t_w\to\infty$ \cite{Fuchs08,Henrich07}. Note that in Ref.~\cite{Fuchs08},
only this term for the difference of the correlators is considered.
It vanishes for tagged-particle density fluctuations, since with
$\delta f=\exp[i\vct q\cdot \vct r_s]$, the average of $\delta f^*\delta f$ is unity in any ensemble.

The second term contains a more complicated dependence on both $t_w$ and $\tau$,
and cannot easily be evaluated. But one recognizes that inserting a
projector onto $\sigma_{xy}$ before the $\delta f^*$ term allows to factorize
the integral according to the different time dependences.
The right-hand part containing ${\mathcal U}(\tau-\tau')$ then becomes proportional
to $\langle\sigma_{xy}\delta f^*{\mathcal U}(\tau-\tau')\delta f\rangle$, for
which the operator identity that led to Eq.~\eqref{eq:ZM} can be rolled back
by noting $\langle\sigma_{xy}\delta f^*\delta f\rangle=0$ due to symmetry
($\delta f^*\delta f$ is symmetric in coordinates $x$ and $y$, while
$\sigma_{xy}$ is antisymmetric). Thus, assuming the dominant part of the
last integral in Eq.~\eqref{eq:ZM} to be given by the projection of
$\delta f$ onto $\sigma_{xy}$, we get
\begin{multline}\label{eq:ctapprox}
  C^f_{t_w}(\tau)\approx \alpha_f(t_w)C_0^f(\tau)
  +\dot\gamma\tilde\sigma(t_w)
  \left\langle\sigma_{xy}\delta f^*e^{\smol^\dagger \tau}\delta f\right\rangle
  \,,
\end{multline}
where we have abbreviated the static renormalization by
$\alpha_f(t_w)=1+\dot\gamma\int_0^{t_w}ds\,\langle\sigma_{xy}\exp[\smol^\dagger s]
\delta f^*\delta f\rangle/\langle\delta f^*\delta f\rangle$. In Refs.~\cite{Krueger09, Krueger10} the simplified version with $\alpha_f(t_w)=1$ was considered which, again, holds exactly for tagged particle dynamics.

Note that formally, we had to introduce a projector
$\proj_\sigma=\sigma_{xy}\rangle\langle\sigma_{xy}\sigma_{xy}\rangle^{-1}
\langle\sigma_{xy}$ that is ill-defined for the case of hard spheres, as there,
the instantaneous shear modulus $\langle\sigma_{xy}\sigma_{xy}\rangle$
diverges \cite{LionbergerRussel,Fuchs03}. However, we only require
\begin{equation}\label{eq:shmod}
  \tilde\sigma(t_w)=\int_0^{t_w}\frac{\langle\sigma_{xy}e^{\smol^\dagger s}
  \sigma_{xy}\rangle}{\langle\sigma_{xy}\sigma_{xy}\rangle}\,ds\,,
\end{equation}
i.e., the integrated normalized shear modulus \cite{Fuchs05,Fuchs05b,
Fuchs06,Crassous08,Hajnal08}, to exist. We assume that this integral
can be regularized for hard spheres, as outlined in Appendix~\ref{app:hs}.

The remaining correlation function in Eq.~\eqref{eq:ctapprox}
is nothing but
the \emph{waiting-time derivative} of $C_{t_w}^f(\tau)$ at $t_w=0$
\cite{Krueger09}, as is immediately clear from taking the $t_w$-derivative
on both sides of Eq.~\eqref{eq:2ti},
\begin{equation}
\dot\gamma\left\langle\sigma_{xy}\delta f^*e^{\smol^\dagger \tau}\delta f
\right\rangle=\left.\frac{\partial}{\partial t_w} C_{t_w}^f(\tau)\right|_{t_w=0}\,.
\label{eq:chi2}
\end{equation}
It describes the initial change of the two-time correlator with $t_w$
at fixed correlation-time window $\tau$. Our approximation then reads
\begin{equation}\label{eq:twotime}
  C^f_{t_w}(\tau)\approx\alpha_f(t_w) C^f_0(\tau)
  +\tilde\sigma(t_w)
  \left.\frac{\partial}{\partial t_w} C^f_{t_w}(\tau)\right|_{t_w=0}\,,
\end{equation}
and Eq.~\eqref{eq:ctapprox} can be interpreted as ``coupling at $t_w\to 0$'',
incorporating the exact result to first order in $t_w$: recalling
$\tilde\sigma(t_w)=t_w+{\mathcal O}(t_w^2)$ and
$\alpha(t_w)=1+{\mathcal O}(t_w^2)$,
\begin{equation}
  C^f_{t_w}(\tau)=C^f_0(\tau)+\left.\frac{\partial}{\partial t_w}
  C^f_{t_w}(\tau)\right|_{t_w=0} t_w+\mathcal{O}((\dot\gamma t_w)^2)
  \,.\label{eq:firstorder}
\end{equation}
Equation~\eqref{eq:ctapprox} extends this identity to finite $t_w$ by
accounting for the static change $\alpha_f(t_w)$ exactly (in principle),
and relating the
further $t_w$ dependence to the integrated shear modulus $\tilde\sigma$.

We still have to close this approximation by relating the waiting-time
derivative to known correlation functions. As was shown in
Refs.~\cite{Krueger09, Krueger10}, one can, using integration by parts and the
identity $\dsmol^\dagger\delta f=0$, arrive at
\begin{multline}\label{eq:ableitungen}
  \left.\frac\partial{\partial t_w}C^f_{t_w}(\tau)\right|_{t_w=0}
  =\left\langle\delta f^*\dsmol^\dagger
  e^{\smol^\dagger \tau}\delta f\right\rangle\\
  =\frac{\partial}{\partial \tau}C^f_0(\tau)
  -\left\langle\delta f^*\smol^\dagger_e e^{\smol^\dagger \tau}
  \delta f\right\rangle\,.
\end{multline}
This equation highlights the connection of the waiting-time derivative
to time derivatives of correlation functions: the derivative of the
transient correlator $C^f_0(\tau)$ has two parts, one containing the
equilibrium operator $\smol^\dagger_e$, and one containing the nonequilibrium
shear-induced $\dsmol^\dagger$. The former term corresponds to
the short-time dynamics of the correlation function, unaffected by shear
as long as $\Pe_0\ll1$, while the latter term, the waiting-time derivative,
is governed by the shear-induced decay of the correlator at long times.

The equilibrium derivative $\smol^{\dagger}_e\delta f^*$ in the last term of
Eq.~\eqref{eq:ableitungen} de-correlates quickly as the
particles loose memory of their initial motion even without shear. In this
case, the latter term is the time derivative of the equilibrium correlator,
$C_e^f(\tau)$. A shear flow switched on at $\tau=0$ will generally lead
to even faster decorrelation, prompting us to approximate
$e^{\smol^\dagger \tau}\approx e^{\smol^\dagger_e \tau}\proj_f
e^{-\smol^\dagger_e \tau}e^{\smol^\dagger \tau}$. This approximation used in the last term in Eq.~\eqref{eq:ableitungen} as well as in $C^f_0(\tau)$ leads to
\begin{equation}
  \frac{\left\langle\delta f^*\smol^\dagger_e\exp[\smol^\dagger \tau]\delta f
  \right\rangle}{\left\langle\delta f^*\exp[\smol^\dagger \tau]\delta f
  \right\rangle}
  \approx\frac{\left\langle\delta f^*\smol^\dagger_e\exp[\smol^\dagger_e \tau]
  \delta f\right\rangle}{\left\langle\delta f^*\exp[\smol^\dagger_e \tau]
  \delta f\right\rangle}
\end{equation}
and thus
\begin{equation}
\left\langle\delta f^*\smol^\dagger_ee^{\smol^\dagger \tau}\delta f\right\rangle
  \approx\frac{C_0^f(\tau)}{C_e^f(\tau)}\frac\partial{\partial \tau}
  C_e^f(\tau)\,.\label{eq:fast}
\end{equation}
Inserting this approximation in Eq.~\eqref{eq:ableitungen} yields
\cite{Krueger09}
\begin{equation}
  \left.\frac{\partial}{\partial t_w} C^f_{t_w}(\tau)\right|_{t_w=0}
  \approx\frac\partial{\partial \tau}C^f_0(\tau)
  -\frac{C^f_0(\tau)}{C^f_e(\tau)}\frac\partial{\partial \tau}C^f_e(\tau)\,.
  \label{eq:lti}
\end{equation}
The two terms in this equation have an intuitive interpretation:
if $\dot\gamma \tau\ll1$, there holds
$C^f_0(\tau)=C^f_e(\tau)+{\mathcal O}(\dot\gamma \tau)$ \cite{Fuchs03}, and the right hand 
side of Eq.~\eqref{eq:lti} cancels in leading order in $\dot\gamma \tau$.
This is expected on physical grounds, since the short-time decay of the
correlation function is independent of $t_w$ at least for small $t_w$.
For $\tau=0$, Eq.~\eqref{eq:lti} yields zero exactly, in agreement with
Eq.~\eqref{eq:chi2}, where $\langle\sigma_{xy}\delta f^*\delta f\rangle=0$
due to symmetry.
On the other hand, for $\dot\gamma \tau=\mathcal{O}(1)$ with $\Pe\gg1$ (i.e.,
the relaxation time of $C_e(\tau)$ is much larger than the
shear-induced relaxation time ${\mathcal O}(1/\dot\gamma)$),
the last term in Eq.~\eqref{eq:lti} vanishes,
and the waiting-time derivative is given by the time derivative of the transient
correlator. We thus refer to the term on the left hand side and the last term in Eq.~\eqref{eq:lti} as
\emph{long-time} and \emph{short-time} derivatives, respectively.

The approximation leading to Eq.~\eqref{eq:lti} can further be made plausible
by considering states that are glassy in the quiescent equilibrium;
setting the second term on the right-hand side to zero, and writing out the
derivatives, one gets
\begin{align}
  C^f_{\delta t}(\tau)&\approx C^f_0(\tau+\delta t)\,, & &\tau\to\infty\,,
\end{align}
for small $\delta t$. This embodies the physical argument that at large times,
whenever the equilibrium dynamics is frozen and the transient correlator is on the plateau, shear effects set in as function of $t$ rather than $\tau$.

\subsection{Relation for the Two-Time Correlator}\label{sec:mainresult}

Equations \eqref{eq:lti} and \eqref{eq:twotime} taken together yield an
approximation that allows us to study the waiting-time dependence of the
non-equilibrium two-time correlation function,
\begin{multline}
  C^f_{t_w}(\tau)\approx C^f_0(\tau)\left[\alpha_f(t_w)
  \vphantom{\frac\partial{\partial \tau}}\right.\\ \left.
  +\tilde\sigma(t_w)\frac d{d\tau}
  \left(\ln|C^f_0(\tau)|-\ln|C^f_e(\tau)|\right)
  \right]\label{eq:2tifin}
\end{multline}
A brief discussion of this result might be in order.
First, we recognize that for weak shear, $\Pe\ll1$, the second term
in Eq.~\eqref{eq:2tifin} does not contribute, as in this regime
$C^f_0(\tau)\approx C^f_e(\tau)$, and hence the normalized waiting-time dependent
correlation function likewise does not change. This ensures that we correctly
recover linear response. On the other hand, for $\Pe\gg1$, the derivative
of the transient correlation function will dominate the second term
in the equation. 

Let us also note the equivalent approximation to Eq.~\eqref{eq:2tifin}
for the mean-squared displacement, easily derived from the $q\to0$ limit
of the corresponding tagged-particle density correlation function.
Considering the $y$- or $z$-direction for simplicity,
\begin{equation}
  \delta z^2(\tau)=\lim_{q\to0}\frac{1-C^f(\tau)}{q^2}\,,
\end{equation}
where $\delta f=\exp[iqz_s]$ with $z_s$ the $z$-coordinate of the tagged particle. 
Performing the $q\to0$ limit in Eq.~\eqref{eq:2tifin} directly yields
(recall that $\alpha_f(t_w)\equiv1$ for tagged-particle density fluctuations)
\begin{multline}
  \delta z^2_{t_w}(\tau)\approx \delta z^2_0(\tau)
  +\tilde\sigma(t_w)\frac d{d\tau}
  \left(\delta z^2_0(\tau)-\delta z^2_e(\tau)\right)\,.
  \label{eq:statMSD}
\end{multline}
Considering only directions perpendicular to the shear direction, the MSD
is linear in time for long times, i.e., $\delta z^2(\tau)\sim \tau$ as $\tau\to\infty$.
If $\dot\gamma \tau\gg1$ (and thus $\dot\gamma t\gg1$), it is plausible that the transient MSD describes
the same diffusivity as the stationary one, and the time-derivatives
of the two functions have to be equal in that limit, as reproduced by
Eq.~\eqref{eq:statMSD}.

Equations~\eqref{eq:2tifin} and \eqref{eq:statMSD} constitute our main
theoretical result: calculating, e.g., $C^f_0(\tau)$ and $C^f_e(\tau)$ within
MCT, the above equations give access to the general waiting-time dependent
$C^f_{t_w}(\tau)$, including the steady-state correlation function usually
measured in experiments.

The equations can also be rewritten in order
to determine the transient correlation function from the more commonplace
equilibrium and steady-state ones. To this end, note that
Eq.~\eqref{eq:2tifin} is solved by
\begin{multline}\label{eq:inverse}
  C^f_0(\tau)=e^{-\alpha_f(t_w)\tau/\tilde\sigma(t_w)}C^f_e(\tau)\\
  +C^f_e(\tau)\int_0^\tau \frac{e^{-\alpha_f(t_w)(\tau-s)/\tilde\sigma(t_w)}}
  {\tilde\sigma(t_w)}\frac{C^f_{t_w}(s)}{C^f_e(s)}\,ds\,.
\end{multline}
Taking $t_w\to\infty$ yields an expression determining the transient
correlator $C^f_0(\tau)$ in terms of both the equilibrium one,
$C^f_e(\tau)$, and the stationary one, $C^f_\infty(\tau)$.
For small waiting times, $\tilde\sigma(t_w)= t_w+\mathcal{O}(t_w^2)$
can be used, and for tagged-particle correlation functions where
$\alpha_f(t_w)=1$, no unknown parameters remain in Eq.~\eqref{eq:inverse}.
For the mean-squared displacement, a similar transform holds,
\begin{multline}\label{eq:inversemsd}
  \delta z^2_0(\tau)=\delta z^2_e(\tau)\\
  +\int_0^\tau e^{-(\tau-s)/\tilde\sigma(t_w)}
  \frac{\delta z^2_{t_w}(s)-\delta z^2_e(s)}
  {\tilde\sigma(t_w)}\,ds\,.
\end{multline}
Incidentally, these forms are also, for direct testing with computer-simulation
data, more stable numerically than Eqs.~\eqref{eq:2tifin} and \eqref{eq:statMSD}
as they do not involve time-derivatives.
We will therefore use them primarily in Sec.~\ref{sec:compsim} where we
use $C^f_e(\tau)$ and $C^f_{t_w}(\tau)$ taken from computer-simulation data to
assess the quality of the approximation by comparing the calculated
and simulated $C^f_0(\tau)$.

\section{A Schematic MCT Model}\label{sec:schematic}

Equations \eqref{eq:2tifin} and \eqref{eq:statMSD} represent the central
theoretical result of our paper. In this section, we visualize their
physical content by choosing a simple toy model for calculating the transient
two-time correlation functions. For the purpose of our discussion, it is
easiest to choose a schematic model of mode-coupling theory, as these models
are very successful both in analyzing real-world data and for
understanding the generic features of colloidal systems under shear.

\subsection{Schematic Equations}

Let us consider a single, normalized transient correlation function
$C_0(\tau)=\phi(\tau)$, to represent the collective density-fluctuation
correlators for some dominant (nearest-neighbor) length scale.
Recently, a schematic model allowing to treat arbitrarily
time-dependent flow has been proposed \cite{pnas}, whose equations of
motion in the case of steady simple shear reduce to 
\begin{subequations}\label{eq:schematic}
\begin{gather}
  0 = \tau_0\partial_\tau\phi(\tau)+\phi(\tau)
  + \int_0^\tau d\tau' m(\tau,\tau-\tau')\partial_{\tau'}\phi(\tau')\,d\tau' \,, \\
  m(a,b) = h_1(a)h_2(b)\hat m(b)\,,\\
  \hat m(a) = \left[v_1\phi(a)+v_2\phi(a)^2\label{eq:mtau}
  \right]\,,\\
  h_{1,2}(a)=1/\left[1+(\dot\gamma a/\gamma_c)^2\right]\,.
\end{gather}
\end{subequations}
Here, the $h_i(a)$ are \textit{ad-hoc} forms for the strain-induced reduction
of the memory kernel, inspired by the appearance of similarly time dependent
terms in the
original MCT vertices due to wave-vector advection. Its precise form is
not crucial, and we follow the choice of Ref.~\cite{pnas} by choosing a
simple decaying function that is even in the strain. There, $\gamma_c=0.1$
was introduced to model
the typical cage-breaking length scale: strains of about $10\%$
mark the point where noticeable strain reduction of memory effects sets in.
$1/\tau_0$ is an initial decay rate that serves to set the unit
of time.

For $\dot\gamma\equiv0$, Eqs.~\eqref{eq:schematic} reduce to the
well-known $\text{F}_\text{12}$ model of quiescent MCT \cite{Goetze84},
whose solutions provide the equilibrium correlator $\phi_e(\tau)$.
Setting $h_1\equiv1$ further reduces our model to the
$\text{F}_\text{12}^{(\dot\gamma)}$ model \cite{Fuchs03}, originally
proposed for the analysis of steady shear flows. This model provides
excellent fits to the flow curves from large scale simulations
\cite{Varnik06a}, and the extension setting $h_1\equiv h_2$ does not
qualitatively change these flow curves, while keeping a closer connection
to the more general time-dependent flows \cite{pnas}.

The $\text{F}_\text{12}$ model has glass transitions along a line of
coupling parameters $(v_1^c,v_2^c)$, where the long time limit
$f=\lim_{\tau\to\infty}\phi_e(\tau)$ jumps discontinuously from zero to its
critical value $f^c$. The choice $v_2^c=2$ is known to
yield good agreement with the asymptotic features expected for the
hard-sphere glass transition; it implies $v_1^c=2(\sqrt2-1)$.
The separation parameter $\varepsilon$ then serves to quantify the distance
to the transition: we set $(v_1,v_2)=(v_1^c,v_2^c)(1+\varepsilon)$
such that $\varepsilon>0$
indicates glassy states, $\varepsilon<0$ fluid ones.

In order to evaluate Eq.~\eqref{eq:2tifin}, we further need a schematic-model
version of Eq.~\eqref{eq:shmod} yielding $\tilde\sigma(t_w)$. We let \cite{Krueger09,Krueger10}
\begin{equation}\label{sigmatw}
  \tilde\sigma(t_w)=\frac{\dot\gamma}3\int_0^{t_w}\phi(s)\,ds\,,
\end{equation}
which can be regarded as
a schematic version of the generalized Green-Kubo relation
derived within ITT and the MCT approximation \cite{pnas}, where
we approximate the dynamical shear modulus $G(s)\approx\phi(s)/3$, 
a reasonable approximation in particular at long times. The
factor $1/3$ accounts for the fact that the plateau in the shear modulus
is empirically found to be smaller than that of the correlator.
In fact, Eq.~\eqref{sigmatw} neglects prefactors and an anisotropic
wave-vector integral that appears in the microscopic Green-Kubo relation;
this means also that we have lost the correct description of the stress
overshoot. To include this effect, one would need to evaluate
Eq.~\eqref{sigmatw} microscopically, as done e.g.\ in Ref.~\cite{Zausch08}.
Eq.~\eqref{sigmatw} is thus to be regarded merely as a plausible closure
that incorporates the structure of MCT-ITT that $\tilde\sigma(t_w)$ is
dominated by an integral over the density correlation functions, effectively
cut off by the slow relaxation time of those correlators.

Equations analogous to Eqs.~\eqref{eq:schematic} hold for the schematic
transient tagged-particle correlation function $\phi^s(\tau)$; the only
difference is in the precise form of the mode-coupling kernel
$\hat m^s(a)$, Eq.~\eqref{eq:mtau}. Its microscopic expression for
tagged-particle density fluctuations can be worked out
\cite{Krueger09b}, and will be discussed elsewhere. For our purpose,
we copy the form of the well-known schematic quiescent tagged-particle
model, the so-called Sj\"ogren model, $\hat m^s(a)=v_s\phi(a)\phi^s(a)$. Here,
a coupling coefficient $v_s>0$ appears that describes the strength of
the tagged-particle coupling to the collective density fluctuations. This
parameter plays no qualitative role in the further discussion; we fix it to
$v_s=5$.

The transient MSD (in the neutral or gradient direction)
is the solution of a similar memory equation, cf.\ Refs.~\cite{Fuchs98,Zausch08},
\begin{equation}
\tau_0\delta \bar z_0^2(\tau)+\int_0^\tau\bar m^{s}(\tau-\tau')\delta\bar z_0^2(\tau')\,d\tau'=2 \tau\,,\label{eq:perp}
\end{equation}
where we denote the schematic-model transient
MSD by $\delta\bar z_0^2$ in order to
avoid confusion with its microscopic counterpart.
In principle, its memory kernel will not be identical to the one appearing
in the tagged-particle-correlator equation, but the MCT approximation for
the self-density fluctuations imply that both these memory kernels are
bilinear functionals of $\phi$ and $\phi^s$, so that they can be approximated
as equal on the schematic level.
Note however that $\bar m^s$ must be a single-time function, as is found in the full microscopic derivation \cite{Krueger09b}, in order to recover long-time diffusion. We therefore
set $\bar m^s(a)=h_2(a)\hat m^s(a)$.

\subsection{Results of the Schematic Model}

\begin{figure}
\begin{center}
\includegraphics[width=0.9\linewidth]{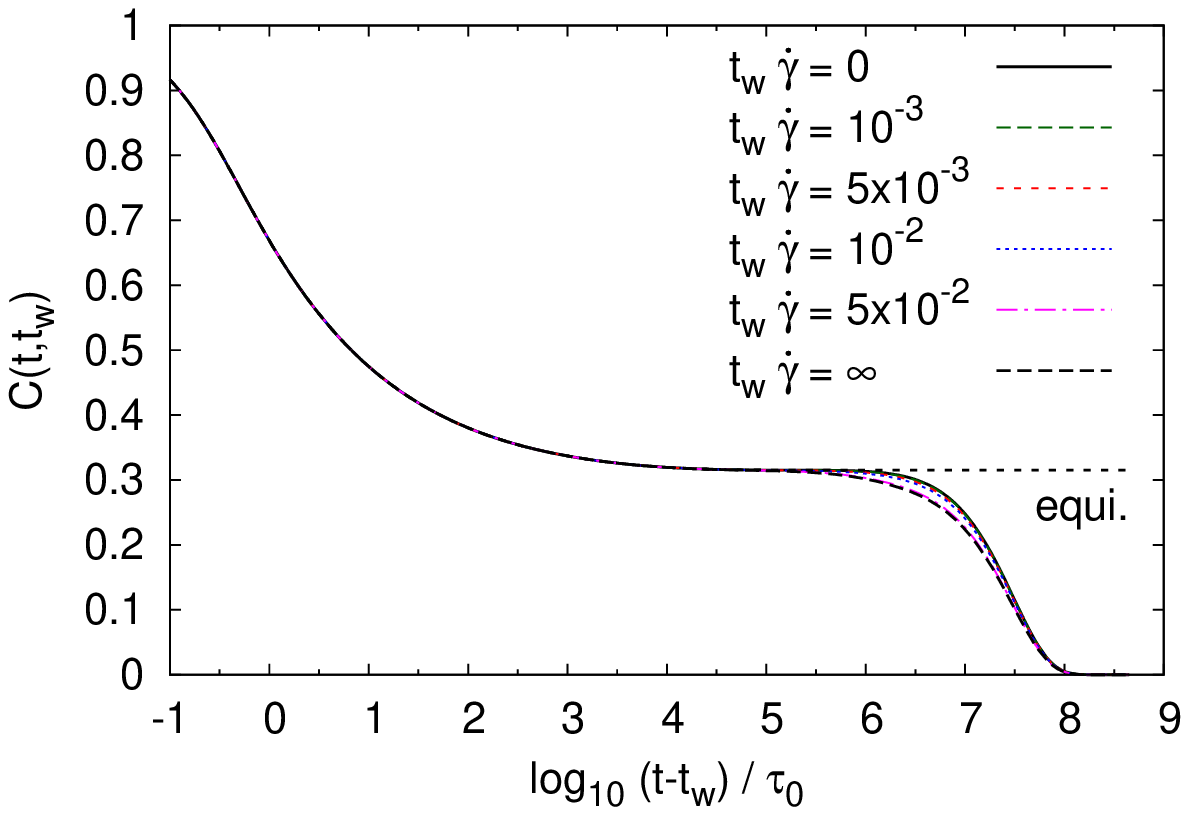}\\
\includegraphics[width=0.9\linewidth]{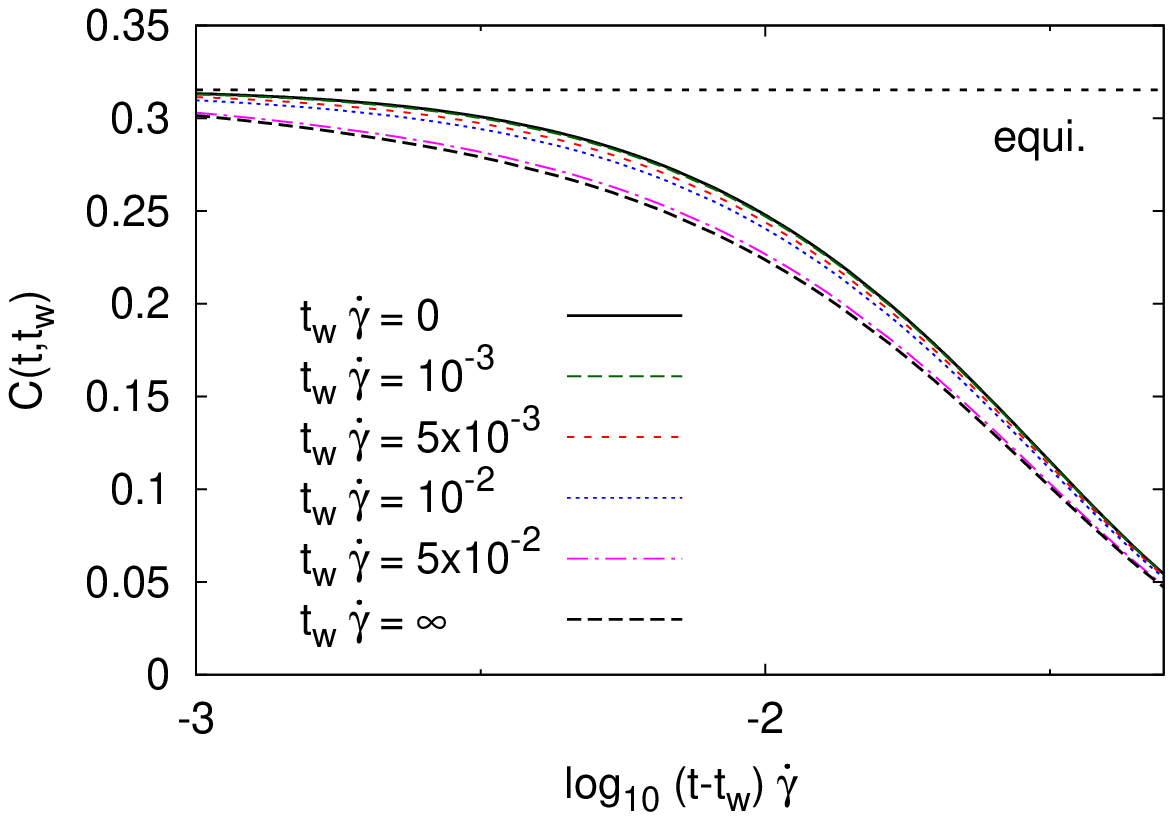}\end{center}
\caption{\label{fig:twotime}
  Dynamical two-time correlation functions,
  Eq.~\eqref{eq:2tifin}, using the $\text{F}_\text{12}^{(\dot\gamma)}$ model \cite{Fuchs03} 
  as input for the transient correlator, in the glass
  ($\varepsilon=10^{-3}$, see text) at shear rate $\dot\gamma\tau_0=10^{-9}$,
  for various waiting times $t_w$ as indicated
  (curves from top to bottom); thick lines indicate the transient ($t_w=0$)
  and steady-state ($t_w=\infty$) correlators. The dotted line represents
  the equilibrium correlation function.
  The lower panel displays the final decay as a function of strain
  $\dot\gamma(t-t_w)$.
} 
\end{figure}

We now turn to a discussion of the central Eqs.~\eqref{eq:2tifin} and
\eqref{eq:statMSD} with the aid of the schematic transient and equilibrium
correlators and MSD defined in the previous section. In order to highlight
the non-trivial effect of the waiting time, we set $\alpha\equiv1$ now,
keeping all correlators normalized to unity at $\tau=t-t_w=0$.
Figure~\ref{fig:twotime}
shows results for $\phi_{t_w}(\tau)$, the schematic waiting-time dependent
correlator calculated via Eq.~\eqref{eq:2tifin} (replacing the general
correlation functions $C_e$ and $C_{t_w}$  with the schematic ones, $\phi_e$ and $\phi_{t_w}$, omitting the $t_w$-subscript only for the transient function $\phi$).
A glassy state was chosen, $\varepsilon=10^{-3}$, so that $\phi_e(\tau)$
attains a finite long-time limit $f$, and the decay of
$\phi_{t_w}(\tau)$ as $\tau\to\infty$ is solely due to the
shear flow. In this case, the last term in Eq.~\eqref{eq:2tifin} vanishes at long times,
and the time derivative of the remaining term is negative for all $\tau$.
The decay of the transient correlator from the plateau can be well
approximated by $\phi(\tau)\approx f\exp[-(\dot\gamma \tau/\gamma_c)^\mu]$
with exponent $\mu=1.2$,
i.e.\ it shows ``compressed exponential'' behavior as a signature
of the non-steady dynamics. Recall that in equilibrium colloidal suspensions,
$\mu>1$ is excluded by the properties of the Smoluchowski operator
\cite{Naegele96,Franosch02}. Approximating the decay for argument's
sake as a simple exponential, we immediately see that
$\phi_{t_w}(\tau)\approx\phi(\tau)(1-\dot\gamma\tilde\sigma(t_w))$ for
$\dot\gamma \tau=\mathcal{O}(1)$.
Since $\tilde\sigma(\infty)$ remains
finite, Eq.~\eqref{eq:2tifin} indeed describes the asymptotic approach
to a steady-state correlator $\phi_\infty(\tau)$ as $t_w\to\infty$.
As seen in the figure, this approach occurs on a time scale
$\dot\gamma t_w\approx5\%$.
The difference between steady-state and transient
correlation functions becomes noticeable only once the correlation functions
decay from their plateau, at $\dot\gamma\tau\approx0.001$;
it vanishes as the functions decay to zero and is most pronounced around
$\dot\gamma\tau\approx0.01$.

\begin{figure}
\begin{center}\includegraphics[width=0.9\linewidth]{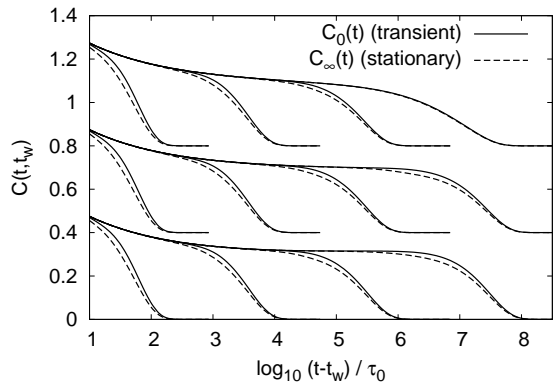}\end{center}
\caption{\label{fig:stat}
  Transient (solid lines) and stationary (dashed lines)
  correlation functions of the schematic model for three state points:
  top curves demonstrate $\varepsilon=-10^{-3}$ (liquid with $\alpha$
  relaxation time $\tau_\alpha/\tau_0={\mathcal O}(10^7)$, shifted by $0.8$
  vertically), middle curves $\varepsilon=0$ (transition point, shifted by
  $0.4$), bottom curves $\varepsilon=10^{-3}$
  (glass). Shear rates are $\Pe_0=\dot\gamma\tau_0=10^{-2n-1}$ with
  $n=1,\ldots 4$ (left to right).
}
\end{figure}

In the liquid, $\varepsilon<0$, similar effects as those described above
are seen, but only in a regime where $\Pe=\dot\gamma\tau_\alpha\gg1$, where
$\tau_\alpha$ is a time scale characterizing the slow relaxation of
the equilibrium correlator $\phi_e(\tau)$ (the ``$\alpha$'' time scale in
glassy liquids), defining the dressed P\'eclet number.
In this case, again, the last term in Eq.~\eqref{eq:2tifin}
can be dropped as a small correction, as $\phi_e(\tau)\approx f$, its plateau
value, for $\tau\approx1/\dot\gamma$. This is visualized in Fig.~\ref{fig:stat},
where results for various $\varepsilon$ and different shear rates $\dot\gamma$
are shown. For $\Pe\gg1$ (and $\Pe_0\ll1$), curves on both sides of
the glass transition show qualitatively identical evolution with $t_w$.
However, in the liquid,
a linear-response regime exists for $\Pe\ll1$, where
Eqs.~\eqref{eq:schematic} describe a transient correlator that is itself
unaffected by shear, $\phi(\tau)\approx\phi_e(\tau)$. This is exemplified
by the top right curve of Fig.~\ref{fig:stat}, where a liquid state of
the $\text{F}_\text{12}$ model with small shear rate corresponding
to $\Pe=\dot\gamma\tau_\alpha=10^{-2}$ was chosen. Equation~\eqref{eq:2tifin}
then correctly describes the fact that all $t_w$-dependent correlation
functions are equal. Note that this holds possibly only
up to a normalization expressed by $\alpha(t_w)$, reflecting the fact that
the static structure may be distorted by shear.

\begin{figure}
\centerline{\includegraphics[width=.8\linewidth]{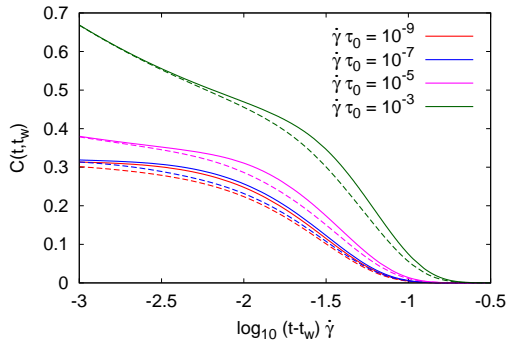}}
\caption{\label{fig:statscal}
  Transient (solid) and stationary (dashed lines) correlators of the
  schematic model in the glass, $\varepsilon=10^{-3}$, for shear rates
  $\Pe_0=10^{-2n-1}$, $n=1,\ldots 4$. 
}
\end{figure}

In the ideal glass, the quiescent relaxation time is infinite, and hence
no linear-response regime exists. Instead, the transient correlator $\phi(\tau)$
decays on a time scale ${\mathcal O}(1/\dot\gamma)$ for arbitrarily small
shear rates. Hence, for large enough
$\dot\gamma \tau$ and $\dot\gamma\to0$, Eq.~\eqref{eq:2tifin} becomes invariant under the
transformation $\tau\mapsto\dot\gamma \tau$ and $t_w\mapsto\dot\gamma t_w$ in the glass.
This yields a nontrivial prediction, namely
that the difference between transient and steady-state correlators
in the shear-molten glass does \emph{not} vanish as $\dot\gamma\to0$.
Rather, a scaling limit is exhibited if one considers the correlation
functions on rescaled times $\dot\gamma \tau$, where both the transient and
the stationary correlators attain (different) master curves.
The approach to this scaling is demonstrated by Fig.~\ref{fig:statscal},
where the $\varepsilon>0$ curves of Fig.~\ref{fig:stat} are reproduced
as functions of $\dot\gamma \tau$. One clearly identifies the approach to two
master
functions looking at the lowest two shear rates, $\dot\gamma\tau_0=10^{-9}$ and
$\dot\gamma\tau_0=10^{-7}$. One notes also that the scaling is
only approached for rather small shear rates, and thus not easily verified.
Although this nonanalytic limit presents an interesting test of MCT,
it will in general be difficult to probe in simulation, as one would need
to construct transient correlation functions in the properly ensemble-averaged
glassy initial state. The physical reason behind this nonanalytic limit
is the existence of a dynamical yield stress predicted by the theory,
rendering the $\dot\gamma\to0$ limit under steady shear a singular one.

Figure~\ref{fig:statscal} demonstrates another preasymptotic effect of the
limit of vanishing shear rate: identifying the regime of accumulated strain
where transient effects are largest, we found $\dot\gamma\tau\approx0.01$
from the small-shear-rate regime discussed above. However, at larger shear
rates, this regime shifts to almost $\dot\gamma\tau\approx0.1$; these are the
$10\%$ strain corresponding to a typical localization length of hard-core
particles in the glass, argued for in Ref.~\cite{Zausch08}.

\begin{figure}
\begin{center}
\includegraphics[width=0.8\linewidth]{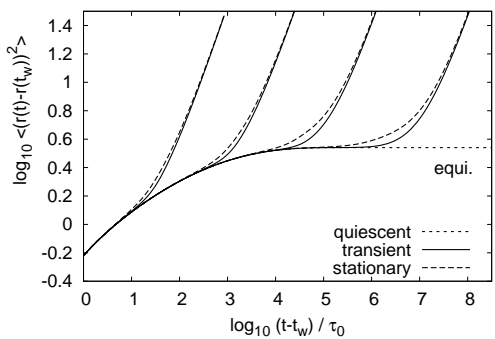}
\end{center}
\caption{\label{fig:statMSD}
  Transient (solid) and stationary (dashed) mean-squared displacements
  in the schematic model, Eq.~\eqref{eq:perp} and Eq.~\eqref{eq:statMSD}, for a glassy state,
  $\varepsilon=10^{-3}$, with shear rates $\Pe_0=\dot\gamma\tau_0=10^{-2n-1}$,
  $n=1,\ldots 4$ (from left to right). The quiescent equilibrium MSD
  is shown as a dotted line.
}
\end{figure}

For the mean-squared displacements, via Eq.~\eqref{eq:statMSD}, a very
similar discussion holds, as exemplified by Fig.~\ref{fig:statMSD}.
The qualitative differences between equilibrium, transient, and steady-state
MSD curves is the same as for the correlators (although now, for obvious
reasons, the functions increase, rather than decrease, with increasing
$t_w$ for fixed $\tau$). The behavior is also in qualitative agreement with
experimental and computer-simulation results of Ref.~\cite{Zausch08}.
However, there, an intriguing super-diffusive regime for the transient
MSD was observed which is missing in our schematic model.
It can be argued that this is due to Eq.~\eqref{eq:perp}, where we model
the memory kernel of the MSD by a strictly positive function $\bar m^s(a)$,
whereas the true memory kernel leading to a superdiffusive regime should
exhibit a small time window of negative values.
Indeed, superdiffusive behavior
was found in Ref.~\cite{Zausch08} to be connected with the stress overshoot
phenomenon (not modeled in our schematic approach)
that was argued to enter $\bar m^s(a)$ via a generalized Stokes-Einstein
approximation.
Thus, the qualitative agreement of the $t_w$-evolution of MSD curves
between our model and the simulation data highlights that the
superdiffusive motion is not necessarily connected to the physics of
crossing over from equilibrium to steady state after switching on shear
flow.

\section{Comparison with Computer Simulation}\label{sec:compsim}

We now turn to a discussion of the waiting time dependence of $C_{t_w}(\tau)$
as found in computer simulation.

We first report the findings from a stochastic-dynamics computer simulation
for a two-dimensional system of hard disks. To avoid crystallization, an
equimolar binary mixture with diameters $\sigma=1$ (taken as the
unit length) and $1.4\sigma$ is chosen; this is the same system
as studied earlier in steady state \cite{fabian_proc}. The number density
is conveniently expressed as the packing (area) fraction of the system
$\varphi$. Previous simulations found a glass transition at packing fractions
around $\varphi\approx0.8$.

The simulation is modeled after the so-called event-driven
Brownian-dynamics (ED-BD) algorithm \cite{edbd}; this algorithm provides
an approximate solution to the stochastic differential equation underlying
Eq.~\eqref{eq:smol} without shear flow,
\begin{equation}
  0=-\zeta dx+dF^\text{r}\,,
\end{equation}
where $x$ is the configuration-space vector describing the particle
positions, $\zeta$ is a friction coefficient
and $F^\text{r}$ is a random white-noise force obeying the
fluctuation-dissipation theorem, $\langle F_i^\text{r}(t)F_j^\text{r}(t')
\rangle=2\kT\zeta\delta(t-t')\delta_{ij}$. Setting the amplitude of the
noise correlation to unity fixes the unit of time.
The interaction among the hard-sphere particles translates into
boundary conditions that no two spheres overlap at any point in time.
The ED-BD algorithm is, in essence, a rejection-free hybrid Monte-Carlo scheme
that works by selecting a small time step $\Delta t$, during which
a free Brownian particle undergoes a displacement with
variance $\langle\Delta x^2\rangle=2D_0\Delta t$ in each Cartesian
direction. Trial moves are first
drawn according to the Gaussian distribution of that variance, and then
corrected for unphysical overlaps. The overlap removal is performed by
assigning to each originally drawn displacement $\Delta x$ a
tangent vector $u=\Delta x/\Delta t$ and a linear curve parameter
$s\in[t,t+\Delta t]$. Particles are then displaced along the tangent vectors
by a linear mapping from $s=t$ to $s=t+\Delta t$.
Whenever at some
$s_c$ two particles $i$ and $j$ start to overlap, the corresponding trial
vectors $u_i$ and $u_j$ are reflected along the plane perpendicular to
the particles' separation vector, which ensures no-flux
boundary conditions on the spheres' surfaces if $\Delta t$ is small enough.
Effectively, this translates
into performing ``elastic collisions'' with the $u_i$ and $u_j$ treated as
velocity vectors. The procedure is continued,
taking care of all of the possibly many $s_c$ in the same fashion until
$s=t+\Delta t$. One is then guaranteed to have a
new configuration that is overlap-free and that the phase space is sampled
ergodically by diffusive motion of a free diffusion coefficient
$D_0=\Delta t/2$.
\footnote{The algorithm has originally been described as assigning
pseudo-velocities to the particles and then performing Newtonian-flight
sub-simulations in every time step of length $\Delta t$.}.

Linear shear flow is incorporated in this algorithm following
Ref.~\cite{fabian_proc} by shifting
the center of the distribution from which displacements are drawn by the known
free-particle drift term. The translation into elastic collisions as in the
flow-free case still ensures ergodicity of the algorithm and can be
expected to be a reasonable approximation for small $\dot\gamma\Delta t$.
Lees-Edwards boundary conditions allow
to match the resulting linear velocity profile with the periodic images
of the simulation box.

For the simulations presented here, we chose a time step
$\Delta t=0.01$,
resulting in $D_0=0.005$.
Initial configurations have been allowed to equilibrate during runs of
up to $D_0t/\sigma^2=4\times10^4$, equivalent to $2\times10^8$ Brownian
time steps. After equilibration, shear flow was instantaneously switched on,
and correlation functions have been measured for several waiting times
$t_w$ thereafter. To improve statistics, the procedure has been repeated
for $300$ independent runs at each density and shear rate.

\begin{figure}
\centerline{\includegraphics[width=.9\linewidth]{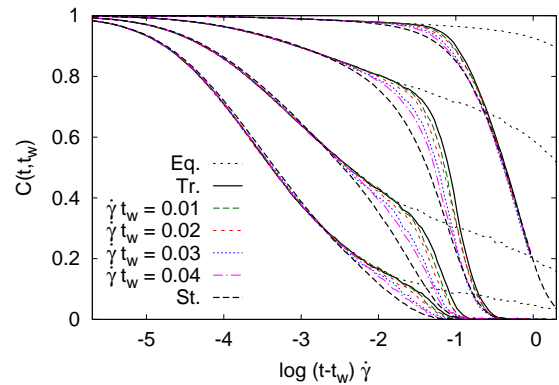}}
\caption{\label{fig:bdsimcorr}
  Tagged-particle density correlation functions from stochastic-dynamics
  computer simulation of a two-dimensional hard-disk system at area fraction
  $\varphi=0.79$, just below the glass transition, and shear rate
  $\Pe_0=0.02$. 
  Curves from top to bottom correspond to wave numbers
  $q\sigma=1.5$, $4.8$, $9.7$, and $14.6$, for wave vectors in the
  direction perpendicular to the flow direction. Thick solid and dashed lines
  are transient ($t_w=0$) and stationary ($t_w\to\infty$) correlators,
  respectively. Thin lines represent different $t_w>0$ as indicated,
  the dotted curves are quiescent equilibrium correlators.
}
\end{figure}

\begin{figure}
\begin{center}\includegraphics[width=0.9\linewidth]{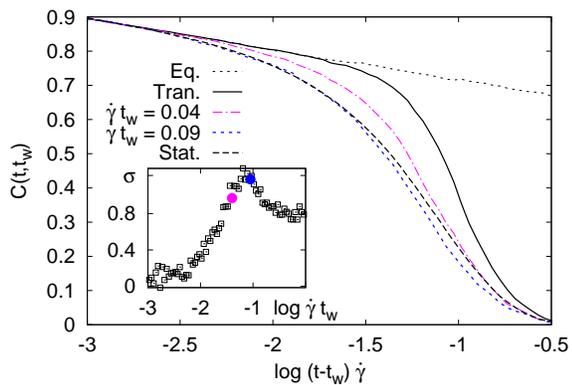}\end{center}
\caption{\label{fig:BDfaster}
  Waiting-time dependent tagged-particle density correlation functions
  for a two-dimensional hard-disk system undergoing stochastic dynamics,
  as in Fig.~\ref{fig:bdsimcorr} for $q\sigma=4.8$,
  but for different waiting times corresponding to accumulated strains
  $\dot\gamma t_w$ where a stress overshoot is seen, marked in the inset.
}
\end{figure}

Figure~\ref{fig:bdsimcorr} shows the self-intermediate scattering functions
for various wave vectors in the gradient direction, at packing fraction
$\varphi=0.79$, and for a fixed shear rate
$\dot\gamma=0.02 D_0/\sigma^2$ ($Pe_0=0.02$). The quiescent correlation
functions are shown for comparison as dashed lines; they decay about two
orders of magnitude slower than the shear-decorrelated ones, hence
$\Pe\approx10^2$.
Different waiting times are shown in dimensionless
units $\dot\gamma t_w$, the scaling expected from our theoretical
observations above.
One recognizes from the figure the same qualitative trends as found in
the Newtonian-dynamics simulation of Ref.~\cite{Zausch08} and consistent
with our theory: the difference
between the various waiting-time dependent correlation functions is most
pronounced at intermediate times, when the functions start to decay
from their respective plateaus.
Following
switch-on, the transient correlation function stays close to its
equilibrium counterpart up to $\dot\gamma \tau\approx{\mathcal O}(0.01)$,
although deviations set in earlier for higher $q$. This is consistent with
the picture that for fluctuations probing smaller length scales, smaller
accumulated strains are needed to deviate from the quiescent state.
Increasing $t_w$, the stationary correlator is approached
for $\dot\gamma t_w\agt0.1$,
somewhat later than in the schematic model discussed above. This results
since in the simulation, also the startup stress $\sigma(t_w)$ approaches
its steady-state value later than in the schematic model (see the discussion
of Fig.~\ref{fig:BDfaster}).
The stationary correlation function deviates earlier
from the equilibrium one than the transient one, and decays slower;
generally all $t_w$-dependent
correlation functions are found to merge again at longer times.
We note in passing that the long-time decay of the transient correlation
function could be fitted with a ``compressed'' exponential function,
resulting in  exponents $\mu=1.1$, $1.8$, $1.8$, and $1.4$ for the four
different wave-vectors shown.
In the corresponding MD simulation of Ref.~\cite{Zausch08},
values of $\mu$ ranging from $1.2$ to $2.4$ have been found, increasing
with increasing wave number \cite{ZauschPhD}.

\begin{figure}
\centerline{\includegraphics[width=.9\linewidth]{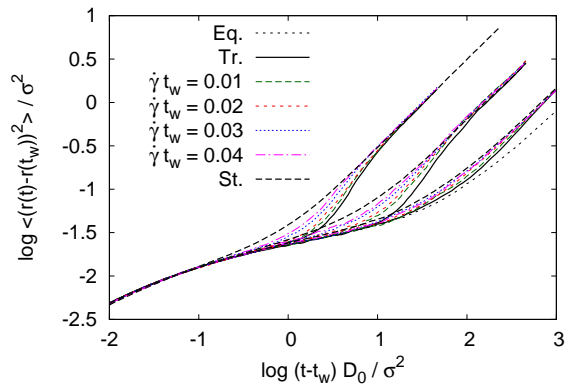}}
\caption{\label{fig:bdsimmsd}
Mean-squared displacements in the shear-gradient direction for the
stochastic-dynamics computer simulation of a hard disk system at area fraction
$\varphi=0.79$, as in Fig.~\ref{fig:bdsimcorr}, but for different Peclet numbers, $Pe_0=2\times10^{-2}$, $2\times10^{-3}$ and $2\times10^{-4}$ from left to right. Dotted line: quiescent equilibrium MSD.
}
\end{figure}

Figure~\ref{fig:bdsimmsd} displays the mean-squared displacements obtained
from the ED-BD simulation, for fixed packing fraction
$\varphi=0.79$, but various shear rates $\dot\gamma$ covering
the regime $\Pe>1$.
Again in qualitative agreement with previous
Newtonian-dynamics results, and also with MSD curves obtained from
confocal microscopy on colloidal suspensions \cite{Zausch08}, for fixed
$\dot\gamma$, the curves for different $t_w$ all collapse for short and
for long times onto the steady-state curve, deviating at intermediate
times; the transient MSD in the simulation
deviates from the equilibrium curve at $\dot\gamma \tau\approx0.02$
and crosses over to the steady-state curve via a super-diffusive regime.
Determining an effective exponent via the logarithmic derivative of the MSD,
$d\log\delta r^2(\tau)/d\log \tau$, for the largest $\dot\gamma$ shown in
Fig.~\ref{fig:bdsimmsd} yields $\delta r^2(\tau)\approx \tau^{1.9}$,
comparable with the exponent found in Ref.~\cite{Zausch08}
for the colloidal suspension, slightly smaller than the one
extracted from the Newtonian dynamics simulation ($\approx2.1$).

The results shown in Figs.~\ref{fig:bdsimcorr} and \ref{fig:bdsimmsd}
are to be compared to the schematic-model results shown in
Figs.~\ref{fig:twotime} and \ref{fig:statMSD}. Regarding the cross-over
from transient to stationary correlation functions, the agreement is indeed
qualitative. This holds despite the fact that, as mentioned above,
the schematic model we employed misses the superdiffusive regime in the
MSD for technical reasons.

In the schematic model,
$\tilde\sigma(t_w)>0$ is a monotonically increasing function of $t_w$.
We hence get a sequence of decreasing $C_{t_w}(\tau)$ for increasing $t_w$
shown in Fig.~\ref{fig:twotime}, i.e.,
\begin{align}
  C_{t_{w_1}}(\tau)&\geq C_{t_{w_2}}(\tau)\,, & & \text{for $t_{w_1}\leq t_{w_2}$,}
  \label{eq:grals}
\end{align}
or reverse for the mean-squared displacement. This ordering rule is well
obeyed by the curves shown in Fig.~\ref{fig:bdsimcorr}, and the related
ordering of the MSD is verified in Fig.~\ref{fig:bdsimmsd}.
Note however that in simulations as well as in the microscopic MCT
the startup stress exhibits an overshoot connected to a small negative
dip in the transient dynamical shear modulus \cite{Zausch08,Fuchs03},
rendering $\tilde\sigma(t_w)$ nonmonotonic as a function of $t_w$.
Hence the ordering of the correlation functions given in Eq.~\eqref{eq:grals}
could in principle be violated for a small $t_w$-window, compare Eq.~\eqref{eq:2tifin}. In particular,
$C_{t_w}(\tau)$ for some fixed $t_w={\mathcal O}(0.1/\dot\gamma)$ could
conceivably be smaller than the steady-state correlator. In the schematic
model, this effect is not contained.

However, as shown by Fig.~\ref{fig:BDfaster}, our simulation indeed
indicates such a crossing of correlators as a function of $t_w$. Here,
waiting times were chosen to sample the $\dot\gamma t_w$ region around the
stress overshoot. For $t_w$ corresponding to the maximum stress, $C_{t_w}(\tau)$ is found to
fall below the stationary correlator, while
this is not the case for smaller $t_w$. 
While lending credibility to our approximations, Fig.~\ref{fig:BDfaster}
also indicates that the approximation in Eq.~\eqref{eq:2tifin} of factorizing $\tilde\sigma(t_w)$  is an oversimplification. For example at $\dot\gamma t_w\approx0.04$, the startup stress measured in the
simulation reaches its steady-state value for the first time, before
entering the overshoot region, and still, the corresponding correlator
$C_{t_w}(\tau)$ differs from the stationary $C_\infty(\tau)$.

The absence of the stress overshoot in our schematic model also implies
that the stationary stress is reached earlier; waiting times of the order
$\dot\gamma t_w\approx0.1$ were sufficient to enter the stationary regime
in Fig.~\ref{fig:twotime}. In the simulation, the corresponding $t_w$ are
slightly larger, since one has to wait for the stress-overshoot region to
be surpassed. The latter causes the transient correlation functions to
approach the steady-state ones only for $\dot\gamma t_w\approx1$.

Motivated by the qualitative agreement,
we now turn to a more general test of Eq.~\eqref{eq:2tifin},
by checking the predicted relation among the three correlation functions
(equilibrium, stationary, and finite $t_w$) for the simulation data.
This will be done both for the Brownian-dynamics data set just discussed,
and also for the Newtonian-dynamics simulation data found in
Ref.~\cite{Zausch08}, in order to test the generality of our approximation
regarding different forms of the short-time dynamics.

In performing the comparison to follow, we are burdened by the fact that
the calculation of $C_{t_w}(\tau)$ from $C_0(\tau)$ and $C_e(\tau)$ performed with
experimental or simulation data is quite unstable, due to the roughening
effect of the numerical derivative and a cancellation of small terms when
all correlators are close to their plateau values. 
In 
the schematic model, $C_0(\tau)$ and $C_e(\tau)$
were available with high enough precision. We therefore have to turn
toward Eq.~\eqref{eq:inverse} and, for the MSD, Eq.~\eqref{eq:inversemsd}.
Although mathematically identical, these forms give fewer numerical
difficulties, as the differentiation can be replaced by a much smoother
numerical integration. Unfortunately, this makes the procedure somewhat
less intuitive: given $C_e(\tau)$ and $C_{t_w}(\tau)$ for some $t_w$, we can now
calculate the transient correlator $C_0(\tau)$ but will, due to the nature of
the approximations involved or due to numerical inaccuracies, get differing
predictions for $C_0(\tau)$ from different $t_w$. This will thus test the accuracy of our approximations for different $t_w$.

In addition, $\tilde\sigma(t_w)$ appearing in Eqs.~\eqref{eq:inverse}
and \eqref{eq:inversemsd} is in general not known or easily determined.
However, for small $t_w$, it can be replaced by its first order expansion, 
and we have a parameter-free prediction for $C_0(\tau)$ out of
$C_e(\tau)$ and $C_{t_w}(\tau)$ that we test in the next section.
After that, we turn to the case $t_w\to\infty$, to demonstrate with
$\tilde\sigma(\infty)$ taken as a (wave-vector independent) fit parameter
the qualitative agreement of our result with the data.

\subsection{Small Waiting Times}

\begin{figure}
\centerline{\includegraphics[width=1\linewidth]{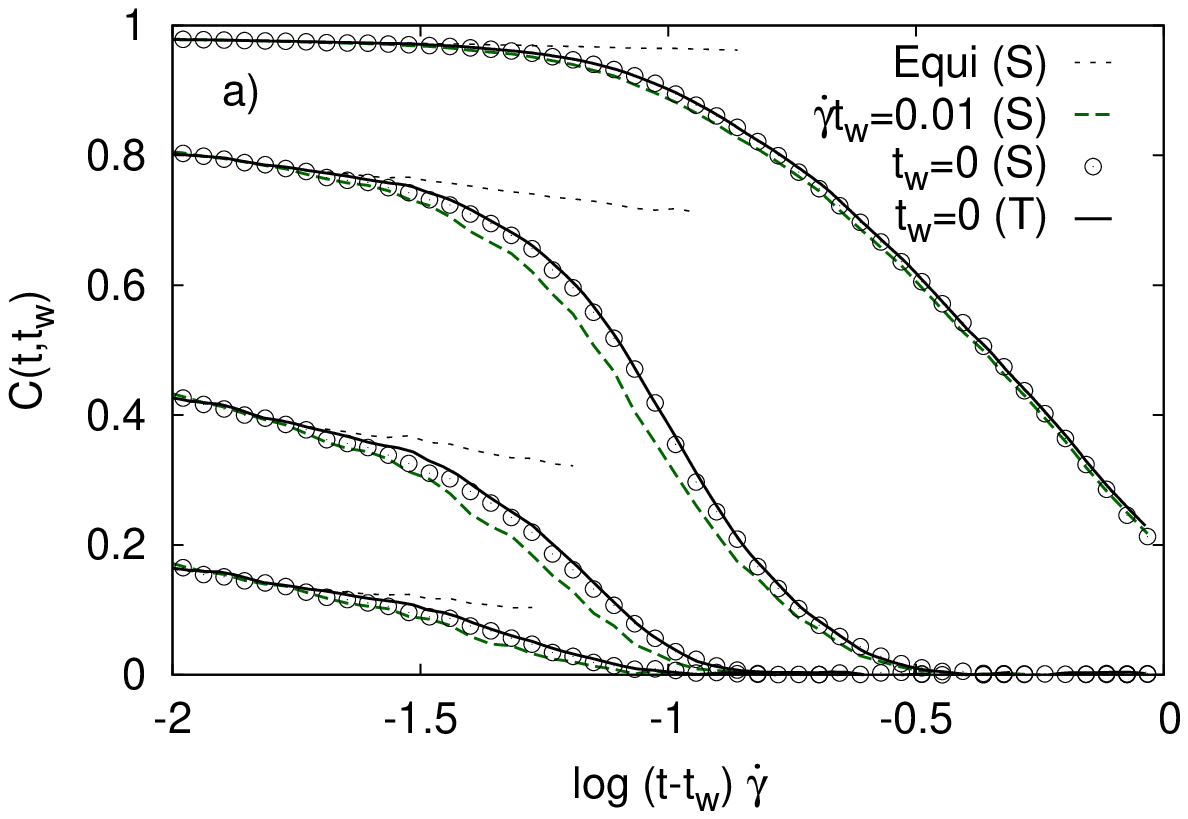}}
\centerline{\includegraphics[width=1\linewidth]{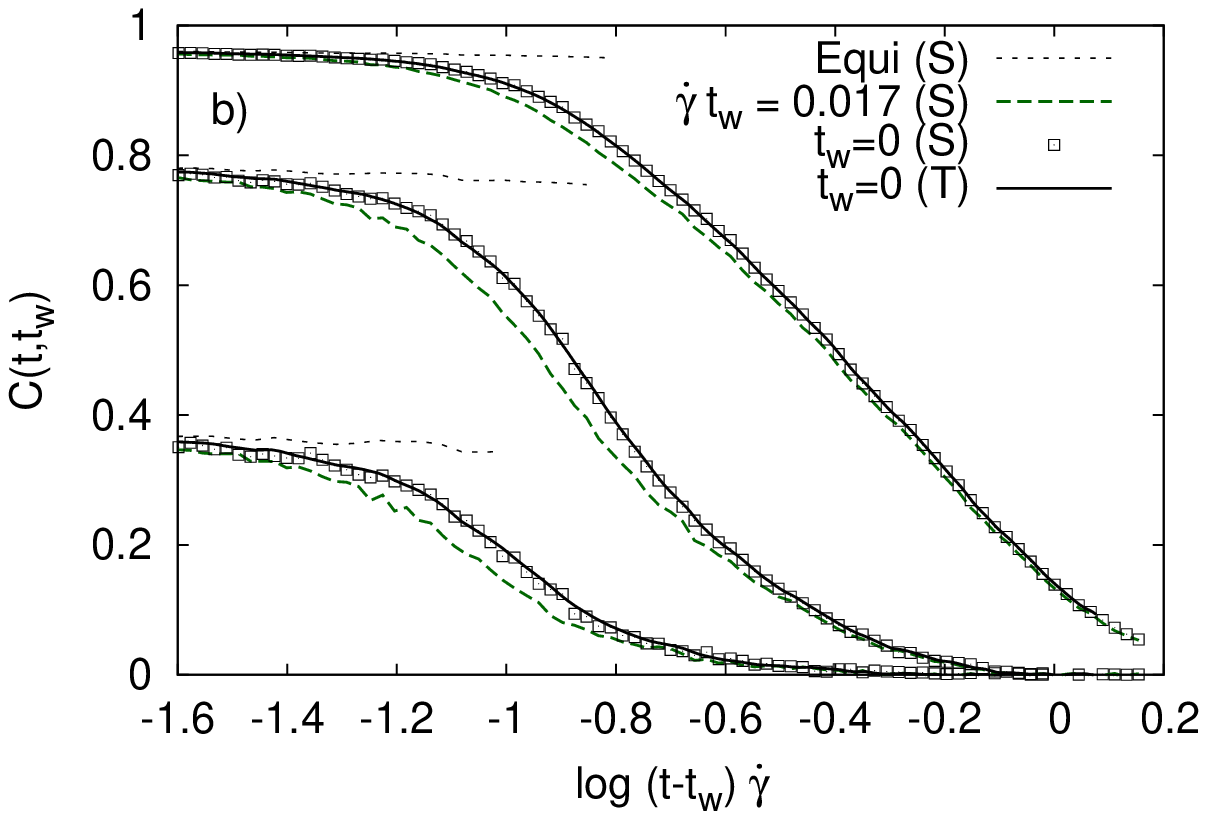}}
\caption{\label{fig:compsimtw}
  (a) Parameter-free
  calculation of the transient correlation function from Eq.~\eqref{eq:inverse}
  from event-driven Brownian-dynamics (ED-BD) computer simulation of hard disks:
  dotted and dashed lines
  are the equilibrium and small-$t_w$ correlation functions,
  respectively, at $\Pe\approx100$, $Pe_0=0.02$ and $\dot\gamma t_w=0.010$,
  with parameters
  as in Fig.~\ref{fig:bdsimcorr}. Solid lines indicate
  the resulting $t_w=0$ transient as calculated from the approximation,
  Eq.~\eqref{eq:inverse}, while symbols show the corresponding
  curves determined from computer simulation directly.
  (b) Same calculation, but for molecular-dynamics (MD) simulation data
  of Ref.~\cite{Zausch08}, for a 3D binary soft-sphere mixture at
  $\Pe\approx10^3$, $\dot\gamma t_w=0.017$,
  and wave vector magnitudes $q\sigma=2.3$,
  $6.0$, and $12.3$ (top to bottom) in the gradient direction.
}
\end{figure}

Figure~\ref{fig:compsimtw} demonstrates the quality of our approximation,
Eq.~\eqref{eq:2tifin}, for small waiting times, where
$\tilde\sigma(t_w)= t_w+\mathcal{O}(t_w^2)$ holds. Both Brownian dynamics
and MD simulation results were used to probe the relation among the three
correlator types via Eq.~\eqref{eq:inverse}. For the ED-BD simulation,
the density is given by $\varphi=0.79$, and $Pe_0=0.02$, as in Fig.~\ref{fig:bdsimcorr}. For the MD, the temperature is given by $T=0.14$ (close to the glass transition on the fluid side) and
$\dot\gamma \sqrt{(m\sigma^2)/\epsilon}=6\times10^{-4}$, where $\epsilon$ sets the particle interaction strength, see Ref.~\cite{Zausch08} for details.
Note that in the figure,
several wave vectors $\vct q$ in the gradient direction were chosen.
Recall that in Eqs.~\eqref{eq:2tifin} and \eqref{eq:inverse}, the
wave-vector dependence is only implicit through that of $C_e(\tau)$ and $C_{t_w}(\tau)$. 
Since the projection onto stresses becomes exact for small waiting times, see Eq.~\eqref{eq:firstorder}, Fig.~\ref{fig:compsimtw} shows explicitely the accuracy of our approximation for the waiting time derivative, Eq.~\eqref{eq:lti}. Noting that this approximation was central for the derivation
 of nontrivial fluctuation-dissipation-ratios in Refs.~\cite{Krueger09,Krueger10} gives strong support for the validity of the results found there:
Fig.~\ref{fig:compsimtw} points out that this approximation 
captures well the observable dependence, at least for the tagged-particle
correlation functions we study.
Let us emphasize again, that for this comparison, no free parameter
appears in the equations, as $\alpha(t_w)=1$ holds for tagged-particle
density fluctuations, and $\tilde\sigma(t_w)$ can be expanded to first
order in $t_w$. Nevertheless, both ED-BD and MD simulation data for the
transient correlation function show excellent agreement with the
curve calculated via Eq.~\eqref{eq:inverse}.

\begin{figure}
\centerline{\includegraphics[width=1\linewidth]{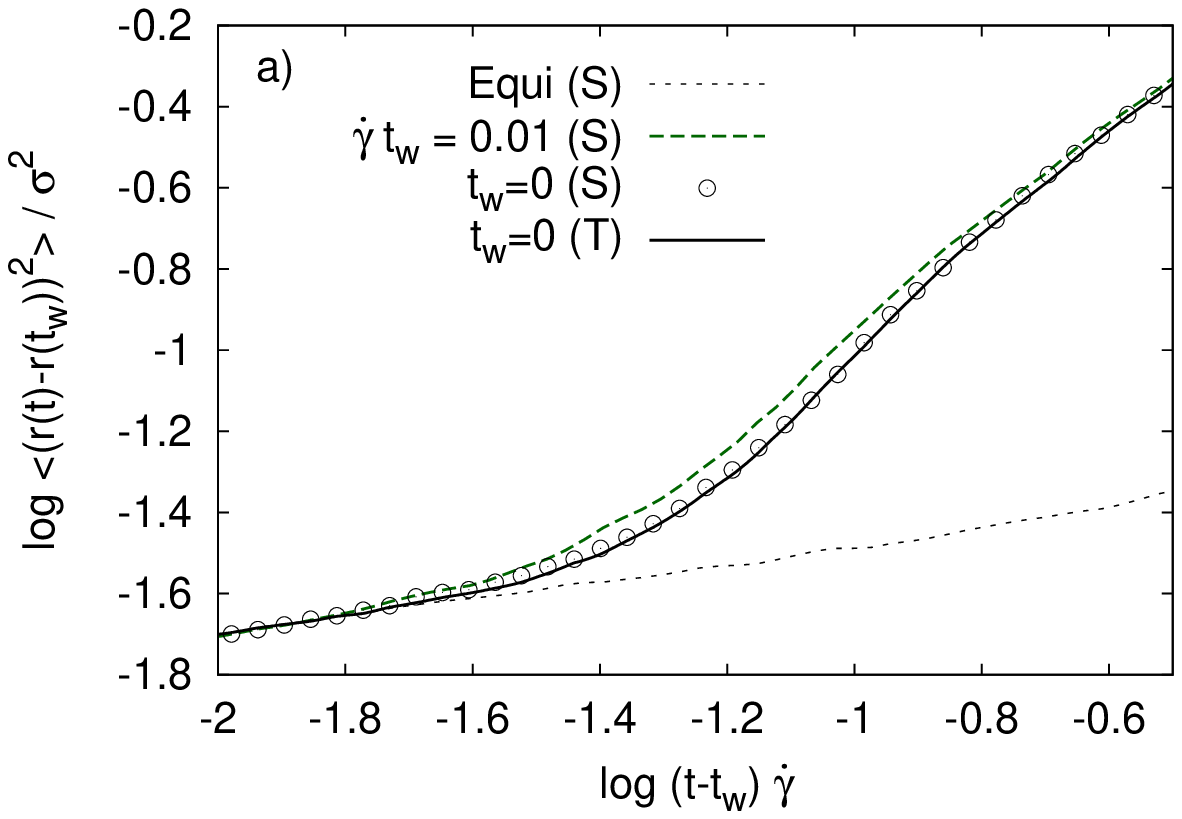}}
\centerline{\includegraphics[width=1\linewidth]{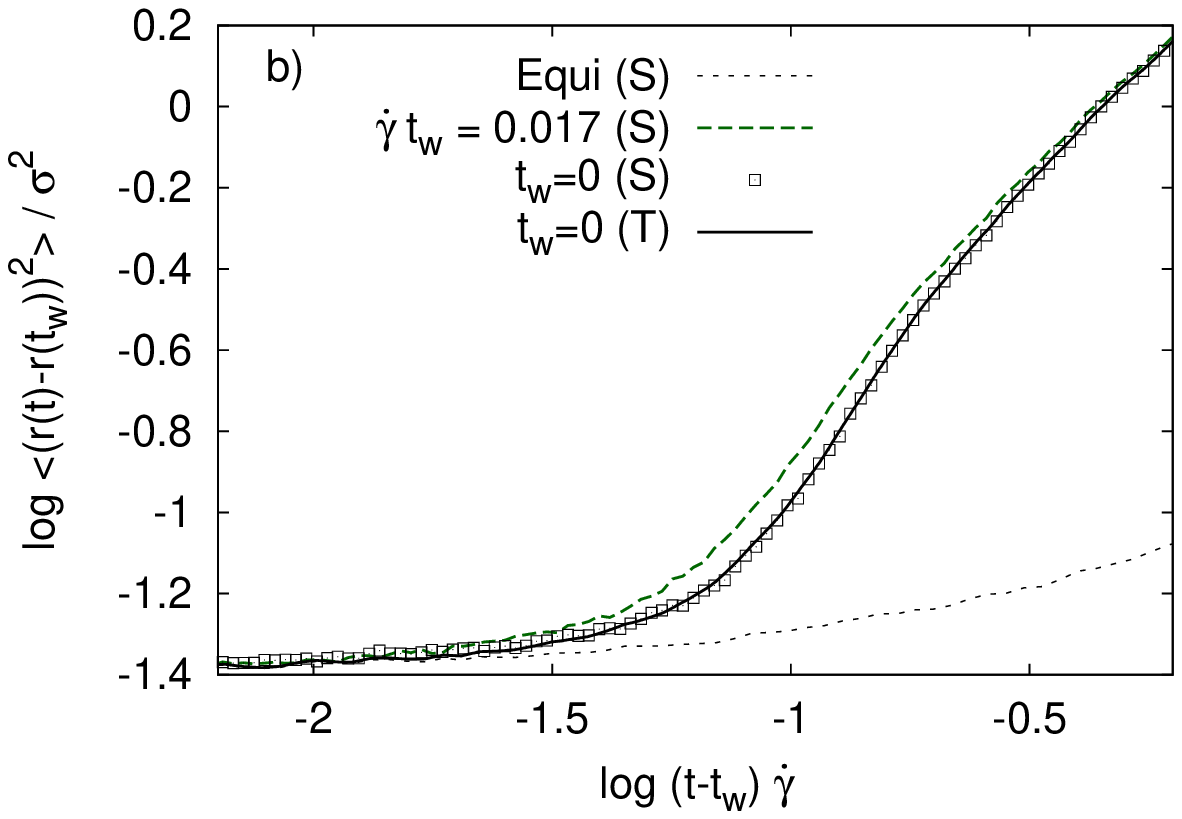}}
\caption{\label{fig:compsimmsdtw}
  Parameter-free
  calculation of transient mean-squared displacements from equilibrium
  and small-$t_w$ data via Eq.~\eqref{eq:inversemsd}, parameters and symbols
  as in Fig.~\ref{fig:compsimtw}.
  (a) ED-BD computer simulation.
  (b) MD computer simulation.
}
\end{figure}

A similarly good agreement is found for the mean-squared displacements
in the directions perpendicular to the shear flow,
as demonstrated by Fig.~\ref{fig:compsimmsdtw}, again
for both Brownian dynamics and MD simulation.
In the latter, to improve statistics, the MSD has been
averaged over both the $y$- and $z$-direction.
The figure in particular highlights that the failure to reproduce superdiffusive
MSDs is completely within the schematic model we chose to illustrate
the equations in Sec.~\ref{sec:schematic}.
Similar to above, the approximation for the waiting time derivative
contained in Eq.~\eqref{eq:statMSD} was central in deriving the
nonequilibrium Einstein relation, Ref.~\cite{Krueger09d}.
The accuracy of this approximation as shown in Fig.~\ref{fig:compsimmsdtw} hence supports the relations found there.

\subsection{Large waiting times}

Having verified that Eqs.~\eqref{eq:2tifin} and \eqref{eq:statMSD} give
a quantitatively correct account for the relation between equilibrium,
waiting-time dependent, and transient correlation function for short waiting times, we now turn
to the $t_w\to\infty$ limit, investigating the relationship between
stationary, equilibrium, and transient dynamics. We again follow the
route outlined above, i.e., we determine the transient correlation functions
and mean-squared displacements from the simulated equilibrium and stationary
ones via Eqs.~\eqref{eq:inverse} and \eqref{eq:inversemsd}, and compare
to the simulated transient correlation function in order to test the
validity of our approach.

To continue we then need $\tilde\sigma(\infty)$ entering our relation;
this quantity could in principle be calculated following its definition
in Eq.~\eqref{eq:shmod}.  In practice, we will use $\tilde\sigma(\infty)$ as a fit parameter,
noting that it should be $q$-independent (in general $f$-independent) to be meaningful.

\begin{figure}
\centerline{\includegraphics[width=1\linewidth]{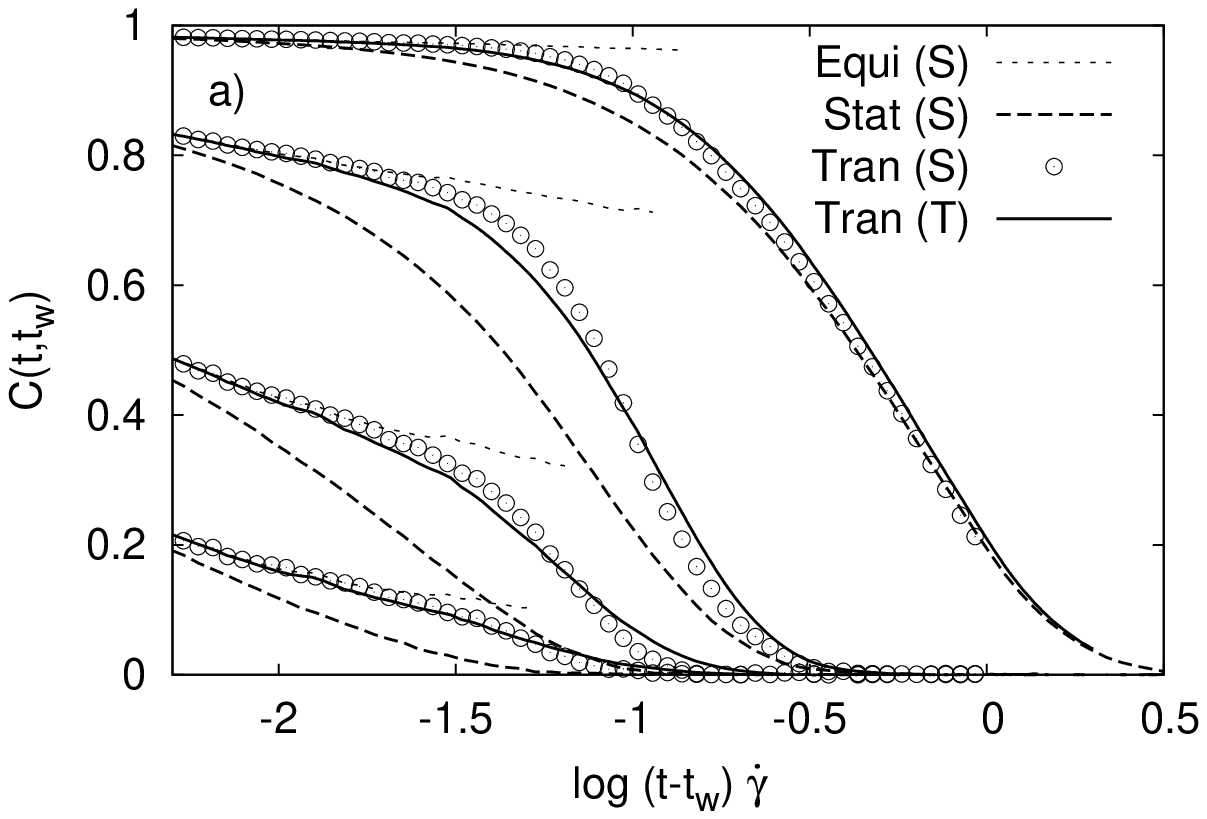}}
\centerline{\includegraphics[width=1\linewidth]{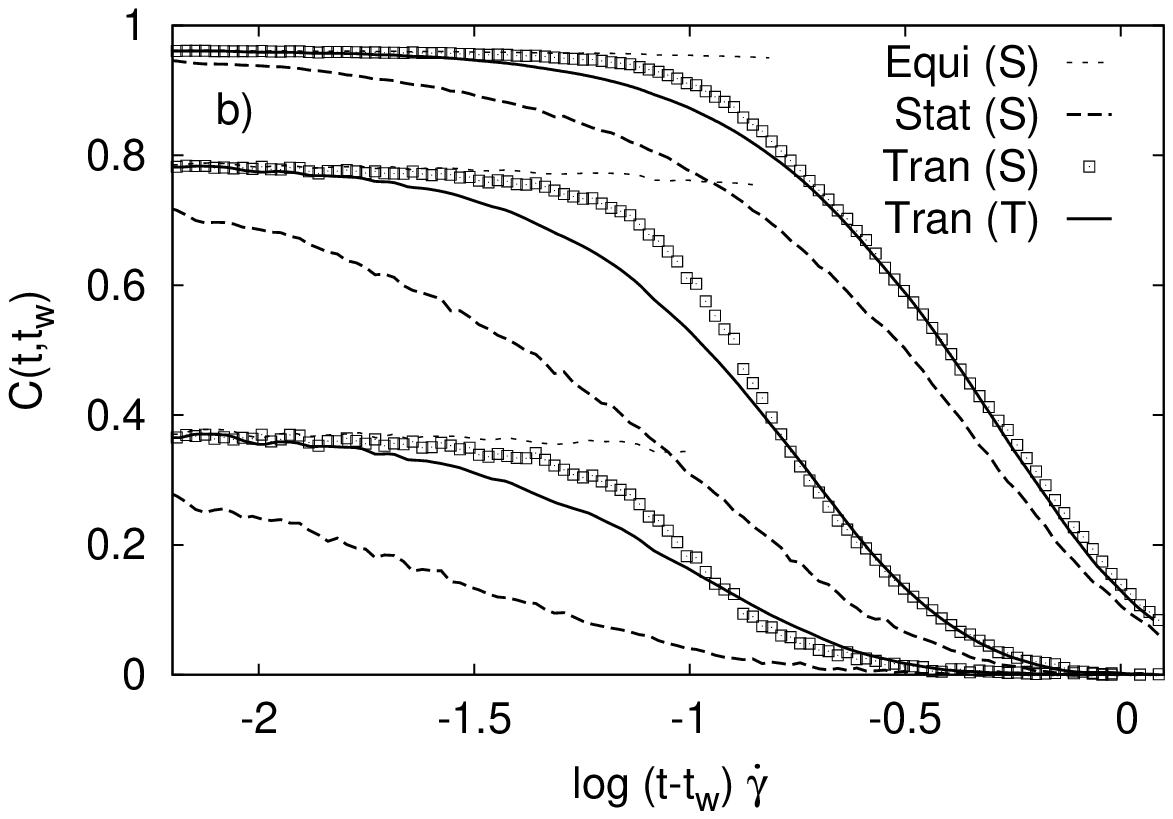}}
\caption{\label{fig:BD}
  Calculation of the transient correlation function from equilibrium
  and steady state via Eq.~\eqref{eq:inverse}. Symbols and parameters
  are as in Fig.~\ref{fig:compsimtw}, but with the small-$t_w$ curve replaced
  by the $t_w\to\infty$ steady-state one. The parameter
  $\tilde\sigma(\infty)$ from Eq.~\eqref{sigmatw} has been fitted.
  (a) Comparison with ED-BD computer simulation data with
  $\dot\gamma\tilde\sigma(\infty)=0.04$.
  (b) Comparison with MD computer simulation data with
  $\dot\gamma\tilde\sigma(\infty)=0.12$.
}
\end{figure}

For the Brownian dynamics data (choosing the same state point and shear
rate as in the small-$t_w$ test above), we find $\dot\gamma\tilde\sigma(\infty)=0.04$ to
give very satisfying results, shown in Fig.~\ref{fig:BD}. For the
molecular-dynamics simulation data, $\dot\gamma\tilde\sigma(\infty)=0.12$ was
fitted differently, as the parameter will depend not only on the shear
rate but also the details of the interaction potential.
Stationary correlators have been obtained from the simulation
after $\dot\gamma t_w=1$, where no significant further change
with $t_w$ was observed.
As expected from the nature of our approximation, which decouples two
time dependences and is hence better for small $t_w$, the agreement for
the simulated and the calculated transient correlator is
still very good, but somewhat less precise than in Fig.~\ref{fig:compsimtw}.
Deviations in the comparison with Brownian dynamics simulation data
are most pronounced for large $q$, where the predicted decay
of the transient correlation function is too slow.

For the molecular-dynamics data, larger deviations occur, as exemplified
in the lower panel of Fig.~\ref{fig:BD}. The same qualitative trend with wave
number
holds as in the comparison with the ED-BD simulation, but the curve shape
predicted for decay of the transient correlation function from the
stationary one is markedly different from the one observed, especially
in the initial decay from the plateau.
Hence, while our approximation seems to hold for both ED-BD and MD equally
well when applied with small enough waiting times, as in
Fig.~\ref{fig:compsimtw}, it is more adapted to ED-BD when applied with
large $t_w$.
Let us point out that
 the time scale for the relaxation of the transient correlator
is captured well in our approximation although it differs
appreciably from the one of the stationary correlator at the larger $q$
shown.
The difference between stationary and transient correlation functions in our theory is proportional to the parameter $\tilde\sigma(\infty)$. Changing it e.g. to smaller values gives transient functions which are closer to the stationary ones in Figs.~\ref{fig:BD} (and also Fig.~\ref{fig:compsimmsd}, see below).

\begin{figure}
\centerline{\includegraphics[width=1\linewidth]{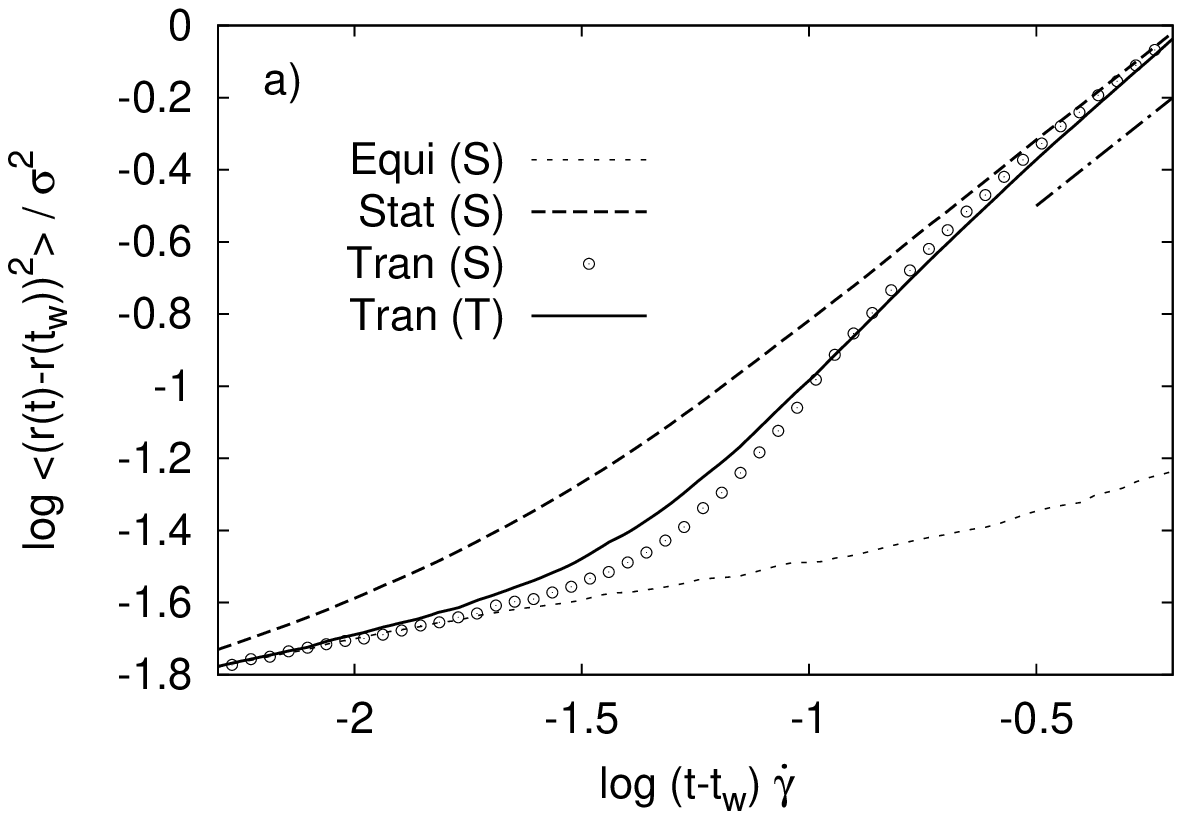}}
\centerline{\includegraphics[width=1\linewidth]{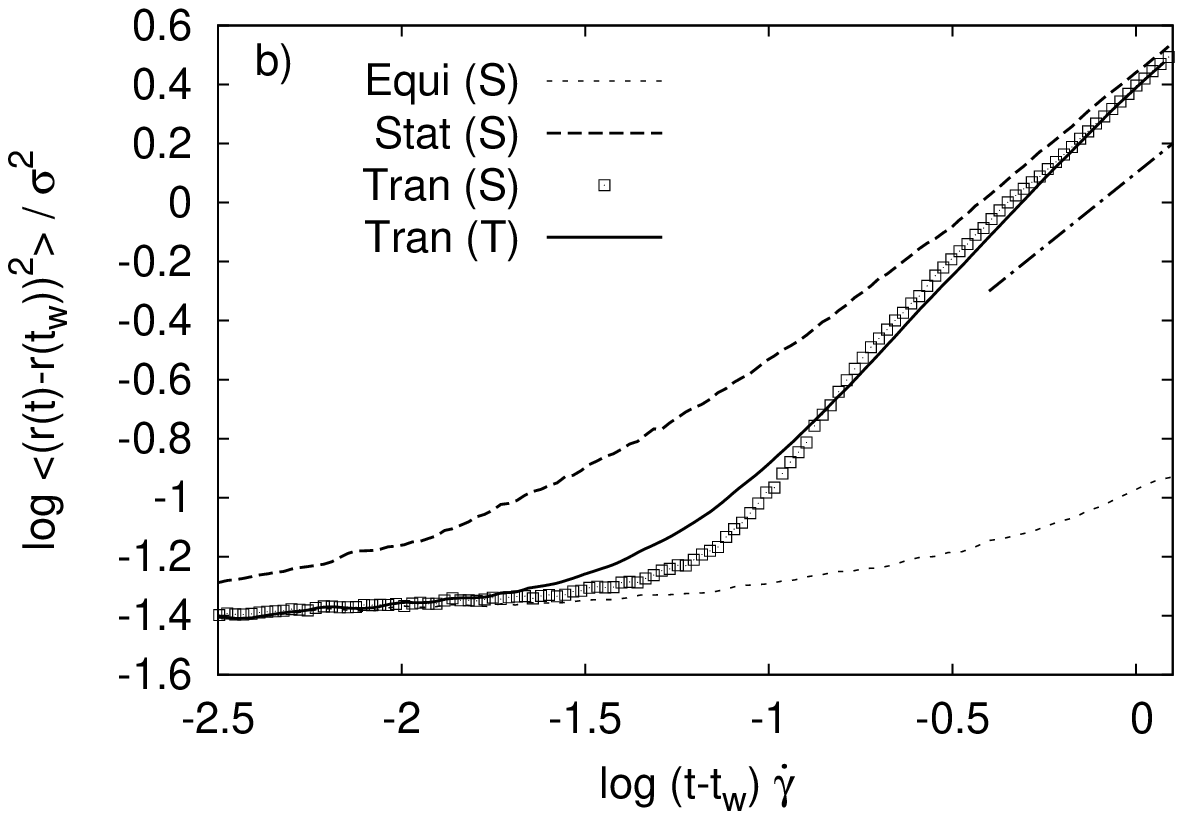}}
\caption{\label{fig:compsimmsd}
  Calculation of transient mean-squared displacements in the gradient
  direction from equilibrium
  and stationary curves via Eq.~\eqref{eq:inversemsd} with $\tilde\sigma(\infty)$-values as in Fig.~\ref{fig:BD}. Curves correspond
  to the parameters as in Fig.~\ref{fig:BD}.
  A dash-dotted line indicates a unit slope corresponding to diffusive motion.
(a) ED-BD computer simulation. (b) MD computer simulation. 
}
\end{figure}

A similar picture arises from the MSD in the gradient direction perpendicular
to the shear flow, shown again for both ED-BD and MD
in Fig.~\ref{fig:compsimmsd}. As in Fig.~\ref{fig:BD}, we have
chosen $\dot\gamma t_w=1$ (ED-BD) and $\dot\gamma t_w\approx0.25$ (MD)
to approximate the stationary MSD in the simulations. The transient MSDs
have been calculated using Eq.~\eqref{eq:inversemsd}, keeping the
fit parameter $\tilde\sigma(\infty)$ fixed as determined from the
comparison of the correlation functions above. The agreement is then seen
to be qualitative again, with some quantitative deviations mostly regarding
the logarithmic slope of the MSD at intermediate times.
These deviations are larger for the MD data than for the 
 ED-BD, and it will be interesting to explore further these
differences in the waiting-time dependent evolution of sheared systems
obeying different types of short-time dynamics.

It appears that, while for the transient dynamics at small $t_w$, possibly
qualitative differences are found between MD and BD (see the discussion above
in conjunction with Fig.~\ref{fig:compsimmsdtw}), this does not seem to be
the case for either the quiescent or the stationary case --
the latter exemplified
by the qualitative similarity of the large-$t_w$ curves in
Fig.~\ref{fig:compsimmsd}.

\section{Discussion}\label{sec:discuss}

We have presented an approximate connection among the different dynamical
two-point correlation functions characterizing the nonequilibrium dynamics
following a sudden commencement of linear shear flow. It describes the
evolution from the pre-shear equilibrium state to the stationary state
attained under shear after sufficiently long waiting times.

The approximations have been derived following the integration-through
transients procedure as a general tool to evaluate averages involving the
unknown nonequilibrium distribution function in terms of history integrals
over the known equilibrium one. ITT is exact in principle, but in evaluating
the two-time three-point correlation functions, we are forced to
introduce approximations. Here we have made
the assumption that the dynamical variables in whose evolution we are
interested in, are, in their evolution from equilibrium to steady state,
dominated by an overlap with the microscopic stress tensor, or more generally
the fluctuations arising from applying the nonequilibrium part of the
time-evolution operator to the distribution function.

An analysis of two different sets of computer simulation data close
to the glass transition
for various wave numbers, shear rates, and
waiting times suggests that our approximative formulas do indeed capture the
essential features of the dynamical evolution from equilibrium to steady state.
To test the generality of the discussed effects with respect to different
short-time dynamics and dimensionality,
we both analyzed previous 3D Newtonian molecular dynamics
data and performed 2D event-driven Brownian dynamics (ED-BD) simulations
leading to diffusive (overdamped stochastic) short-time motion.

We have made plausible the approximations by referring to a schematic
model of mode-coupling theory that allows to calculate all the desired
quantities numerically with high precision. The theoretical predictions
are seen to
broadly agree with previous results on both molecular-dynamics computer
simulations and on colloidal experiments reported in Ref.~\cite{Zausch08},
and also with the hard-sphere stochastic-dynamics simulations we performed.
As a general trend, the transient correlation functions follow the
equilibrium ones for times up to $\dot\gamma \tau\approx0.01$;
after this strain is reached, they deviate and decay much more
rapidly in the regime of high dressed P\'eclet numbers we are interested
in. For the tagged-particle density correlation functions, the deviation sets
in earlier for larger wave number (tested in the gradient direction).
For smaller $\Pe$ (or in the case of the MSD), the decisive strain is closer to
$\dot\gamma \tau\approx0.1$:
as mentioned before \cite{Zausch08}, this value is
suspiciously close to a typical particle localization length, supporting
the picture that the steep decay of the transient correlation function
marks the breaking of nearest-neighbor cages under shear.
As the waiting time $t_w$ increases, this steep decay, found to be
super-exponential in our model, broadens to settle on the
$t_w\to\infty$ stationary limit for waiting times of $\dot\gamma t_w\approx1$.
For most cases, this gives rise to an ordering of $t_w$-dependent correlators,
with the transient, $t_w=0$, correlator being the largest (i.e. slowest),
and the stationary, $t_w=\infty$, correlator being the smallest (i.e. fastest)
in an intermediate window of rescaled times $\dot\gamma\tau$.
However, this ordering is violated for waiting times corresponding to
strains $\dot\gamma t_w$ where a non-monotonic variation is observed in the
shear stress after startup, the so called stress overshoot.
Our approximation recovers this subtlety.

As a side note, it should be mentioned that in the case of switching on
steady shear, the waiting-time dependent correlation functions are found
to all decay on time scales of ${\mathcal O}(1/\dot\gamma)$ when leaving
the linear-response regime, i.e., when $\Pe\gg1$. This is in notable
difference to waiting-time dependences usually discussed in the aging
dynamics of glasses following a sudden temperature quench
or in soft-matter systems
\cite{SUSSP,Cipelletti,Barrat03}, where the final relaxation time strongly
increases as a
function of increasing $t_w$, giving rise to an ordering of correlators
that is the reverse of what we discussed here.

Let us remark that our schematic-model analysis suggests
to discuss the peculiarities of the transient dynamics in colloidal and
non-colloidal systems in separate aspects: for both types of short-time
motion ``compressed exponential'' relaxation is found in simulations,
which in colloidal systems can be seen as the hallmark of nonequilibrium
dynamics (as the equilibrium dynamics is confined to show at most exponential
relaxation for any correlation function as long as only structural relaxation
is involved). In fact, the nonequilibrium
steady-state dynamics according to our observations is again better
described by pure exponentials, so that the compressed
exponentials may be the signature of nonequilibrium and non-stationary
relaxation.
This faster-than-exponential relaxation
does, however, not directly translate into a super-diffusive (or even
ballistic) regime in the mean-squared displacement, another feature observed
in simulations on both types of system. As already investigated in detail
in Ref.~\cite{Zausch08}, this latter feature can arise independently
and is directly related to
the stress overshoot.

The mechanism by which the dynamics evolves from equilibrium via transient
to stationary dynamics appears to have aspects that are independent on the details of the
short-time motion: The points in time where deviations are first
seen from the equilibrium correlator, and last seen with respect to the
steady state, are determined by the total accumulated strain $\dot\gamma \tau$.
The decisive value of $\dot\gamma \tau\approx0.1$ turns out
to be suspiciously close to a typical cage size in supercooled liquids,
as has been noted previously.
We reiterate that this generality in the evolution from equilibrium to
steady state does not mean that an equally strong
statement holds as it does in equilibrium, where the long-time part of
the dynamics itself is independent on the type of short-time dynamics.
In fact, analyzing mean-squared displacements and tagged-particle
density-correlation functions from both ED-BD and MD simulations,
the MD ones appear to show a somewhat stronger transient
effect than the ED-BD ones. Determining effective exponents for the MSD
(from a logarithmic derivative), the MD data gives larger deviations
from sub-diffusive structural relaxation. The
compressed-exponential exponents obtained from the incoherent scattering
function can be significantly larger than $2$ in MD, but remain below $2$
for the ED-BD data we analyzed.
This occurs in a time window, where for the quiescent system structural
relaxation prevails, and where one thus expects universal aspects of slow
relaxation that are independent of the short-time dynamics. It should be
noted that the simulation models are different, so that we cannot exclude
the difference observed in the transients
being due to details of the interparticle interactions.

Our approximation correspondingly shows largest deviations when relating
equilibrium, transient, and stationary correlation functions at large $q$
for the molecular-dynamics simulation, even though it performs surprisingly
well for the same MD data when restricting it to smaller waiting times.
This may indicate that the peculiar form of the $t_w=0$ transient in
molecular dynamics is not fully captured by considering all dynamical
relaxations to be governed by the dynamics of the potential part of the
local stress tensor (as we did to arrive at our approximation), but that
non-potential parts of the dynamics may play a role in this regime, too.

Nevertheless, the overall quality of the waiting-time-derivative approximation
(Eq.~\eqref{eq:lti}) is found to be remarkably accurate. This lends further support to
the earlier discussion of nonequilibrium fluctuation-dissipation ratios
and Einstein relations within MCT and its schematic models, as the
equivalent approximations have been used there
\cite{Krueger09,Krueger09d,Krueger10}.

\begin{acknowledgments}
We wish to acknowledge fruitful and stimulating discussions with
Matthias Fuchs, J\"urgen Horbach, Joseph M. Brader and Jochen Zausch. We thank Jochen Zausch
for providing us with his molecular-dynamics simulation data.
This work was partially supported by the Helmholtz-Gemeinschaft (HGF)
through its Impuls- und Vernetzungsfonds. Th.~V.\ acknowledges
funding through a Helmholtz-University Young Investigator Group, and
through a fellowship of the Zukunftskolleg of the Universit\"at Konstanz.
M.~K.\ acknowledges funding through the International Research Training
Group ``Soft Condensed Matter'', IRTG 667, as well as the
Sonderforschungsbereich Transregio 6
``Physics of Colloidal Dispersions in External Fields''.
\end{acknowledgments}

\begin{appendix}

\section{Hard Spheres} \label{app:hs}

For the hard sphere system, the instantaneous shear modulus diverges,
rendering the projector $\proj_\sigma$ leading to Eqs.~\eqref{eq:ctapprox}
and \eqref{eq:shmod}
ill-defined. For a proper treatment of this singular limit, one can use
instead
\begin{equation}
\proj_\sigma(s_0)=e^{\smol^\dagger s_0}\sigma_{xy}\rangle
  \langle\sigma_{xy}e^{2\smol^\dagger s_0}\sigma_{xy}\rangle^{-1}
  \langle\sigma_{xy}e^{\smol^\dagger s_0}
\end{equation}
for some fixed small $s_0$.
With this choice, Eq.~\eqref{eq:ctapprox} remains unmodified, but the
insertion of
$\exp[\smol^\dagger s_0]\proj_\sigma(s_0)\exp[-\smol^\dagger s_0]$
changes Eq.~\eqref{eq:shmod} to
\begin{equation}
  \tilde\sigma(t_w,s_0)=\int_{2 s_0}^{t_w}
  \frac{\langle\sigma_{xy}e^{\smol^\dagger s}
  \sigma_{xy}\rangle}{\langle\sigma_{xy} e^{2\smol^\dagger s_0}\sigma_{xy}\rangle}\,ds\,,
\end{equation}
where the integrand remains finite for all $t_w\ge2 s_0$.
The approximation thus, in the case of hard spheres, consists of
coupling the variable $f$ to the shear stress at some small cutoff time
$s_0$.

\end{appendix}

\end{document}